\documentclass[lettersize,journal]{IEEEtran}

\usepackage{adjustbox}
\usepackage{algorithm}
\usepackage{amsfonts}
\usepackage{booktabs}
\usepackage{cite}
\usepackage{diagbox}
\usepackage{endnotes}
\usepackage{enumerate}
\usepackage{enumitem}
\usepackage{graphicx}
\usepackage{hyperref}
\usepackage{listings}
\usepackage{longtable}
\usepackage{makecell}
\usepackage{multirow}
\usepackage{nicematrix}
\usepackage{tabularx}
\usepackage{tcolorbox}
\usepackage{xcolor}
\usepackage{xurl}

\usepackage[caption=false,font=normalsize,labelfont=sf,textfont=sf]{subfig}
\usepackage[font=small, skip=2  pt]{caption}

\renewcommand{\floatsep}{5pt}    
\renewcommand{\textfloatsep}{5pt} 
\renewcommand{\intextsep}{5pt}   

\definecolor{red_color}{RGB}{255,0,0}
\definecolor{green_color}{RGB}{0,255,0}
\definecolor{yellow_color}{RGB}{255,255,0}

\tcbuselibrary{skins}

\makeatletter
\newcommand{\ie}{\emph{i.e.}\@ifnextchar.{\!\@gobble}{}}
\newcommand{\eg}{\emph{e.g.}\@ifnextchar.{\!\@gobble}{}}
\newcommand{\etc}{\emph{etc}\@ifnextchar.{}{.\@}}
\newcommand{\aka}{\emph{aka}\@ifnextchar.{}{.\@}}
\makeatother

\newcolumntype{C}[1]{>{\centering\arraybackslash}p{#1}}

\newcommand{\collectdate}{October 12th, 2024}
\newcommand{\interviewdate}{October 22th, 2024}

\newcommand{\defineNewIndex}[1]{
    \newcounter{#1}
    \expandafter\renewcommand\csname the#1\endcsname{\ifnum\value{#1}<10 0\fi\arabic{#1}}
    \expandafter\def\csname use#1\endcsname{%
        \refstepcounter{#1}
        \csname the#1\endcsname
    }
}

\newcommand{\defineOldIndex}[1]{
    \newcounter{#1}
    \expandafter\def\csname use#1\endcsname{%
        \refstepcounter{#1}
        \csname the#1\endcsname
    }
}

\defineOldIndex{workflowIndex}

\newtcolorbox{mybox}[2][]{
top=0.15in,left=4pt,right=4pt,bottom=4pt,
fonttitle=\bfseries,
colbacktitle=gray,
colback=gray!5,
colframe=gray!40!black,
enhanced,
attach boxed title to top left={xshift=0em,yshift=-\tcboxedtitleheight/2},
boxed title style={size=small},
drop shadow={black!50!white},
title=#2,#1}

\begin{document}

\title{On the Workflows and Smells of Leaderboard Operations (LBOps): An Exploratory Study of Foundation Model Leaderboards}

\author{Zhimin~Zhao, Abdul~Ali~Bangash, Filipe~Roseiro~Côgo, Bram~Adams,~\IEEEmembership{Senior~Member,~IEEE}, Ahmed~E.~Hassan,~\IEEEmembership{Fellow,~IEEE}
\thanks{The authors are with the Software Analysis and Intelligence Lab (SAIL), School of Computing, Queen's University, Kingston, ON K7L 3N6, Canada (email: z.zhao@queensu.ca; abdulali.b@queensu.ca; filipe.cogo@gmail.com; bram.adams@queensu.ca; hassan@queensu.ca)}}

\date{Received: date / Accepted: date}

\maketitle

\begin{abstract}
Foundation models (FM), such as large language models (LLMs), which are large-scale machine learning (ML) models, have demonstrated remarkable adaptability in various downstream software engineering (SE) tasks, such as code completion, code understanding, and software development. As a result, FM leaderboards have become essential tools for SE teams to compare and select the best third-party FMs for their specific products and purposes. However, the lack of standardized guidelines for FM evaluation and comparison threatens the transparency of FM leaderboards and limits stakeholders' ability to perform effective FM selection. As a first step towards addressing this challenge, our research focuses on understanding how these FM leaderboards operate in real-world scenarios (``leaderboard operations'') and identifying potential pitfalls and areas for improvement (``leaderboard smells''). In this regard, we collect up to $1,045$ FM leaderboards from five different sources: GitHub, Hugging Face Spaces, Papers With Code, spreadsheet and independent platform, to examine their documentation and engage in direct communication with leaderboard operators to understand their workflows. Through card sorting and negotiated agreement, we identify five distinct workflow patterns and develop a domain model that captures the key components and their interactions within these workflows. We then identify eight unique types of leaderboard smells in LBOps. By mitigating these smells, SE teams can improve transparency, accountability, and collaboration in current LBOps practices, fostering a more robust and responsible ecosystem for FM comparison and selection.
\end{abstract}


\section{Introduction}
\label{sec:introduction}

Foundation models~\cite{bommasani2021opportunities} (FMs), also referred as ``large AI models''~\cite{el2022impossible}, represent a paradigm-shifting advancement in the development of AI-driven software systems. These ML models, characterized by billions of parameters and trained on extensive, diverse datasets, exhibit exceptional flexibility, enabling them to be fine-tuned and adapted for a wide range of downstream tasks, such as code completion~\cite{dakhel2023github,barke2023grounded}, understanding~\cite{nam2024using}, and software development~\cite{qian2023communicative}. With the widespread adoption of model enhancement techniques, such as fine-tuning~\cite{church2021emerging}, knowledge distillation~\cite{gou2021knowledge}, quantization~\cite{gholami2022survey}, instruction tuning~\cite{liu2024visual}, retrieval augmented generation~\cite{lewis2020retrieval} (RAG), prompt engineering~\cite{ziegler2023developer} and agentic workflow~\cite{singh2024enhancing}, selecting the most suitable FMs has become a daunting challenge for software engineering (SE) practitioners~\cite{raschka2018model,zhou2021model}.  

An emerging solution is the use of FM leaderboards: online applications that provide ``ranking-as-a-service'' (RaaS) to evaluate and compare FMs (or FM-powered agents) against a set of ML benchmarks, helping stakeholders make well-informed decisions\footnote{\url{https://huggingface.co/docs/leaderboards}}. Such leaderboards, hosted on different sources, such as \href{https://pages.github.com}{GitHub Pages}, \href{https://huggingface.co/spaces}{Hugging Face (HF) Spaces}, and \href{https://paperswithcode.com/sota}{Papers With Code (PWC)}, facilitate the model selection process by offering structured and objective comparisons of performance between models. These applications are supported by complex operational workflows (\eg, processes to keep the leaderboards functional and reliable) that try to ensure continuous, systematic, and reliable performance comparisons of participating models for diverse stakeholders: software engineers who need reliable model rankings for optimal selection, model producers focused on evaluating their models' performance, and leaderboard operators who aim to maintain and showcase state-of-the-art (SOTA) benchmarks.

However, maintaining a relevant and reliable leaderboard requires significant ongoing effort. On the one hand, decisions regarding benchmark selection, evaluation infrastructure, and workflows for long-term maintenance directly influence the costs and efforts required from leaderboard operators. For example, prominent leaderboards, such as the \href{https://huggingface.co/spaces/open-llm-leaderboard/open_llm_leaderboard}{Open LLM Leaderboard}, engage with thousands of user discussions, many of which center around issues such as \href{https://huggingface.co/spaces/open-llm-leaderboard/open_llm_leaderboard/discussions/854}{failed evaluation}, \href{https://huggingface.co/spaces/open-llm-leaderboard/open_llm_leaderboard/discussions/842}{outdated scores}, \href{https://huggingface.co/spaces/open-llm-leaderboard/open_llm_leaderboard/discussions/856}{unclear documentation}, and \href{https://huggingface.co/spaces/open-llm-leaderboard/open_llm_leaderboard/discussions/540}{incorrectly tagged models}. These issues highlight violations of key software quality attributes—reliability, availability, maintainability, and usability—within the domain of ML leaderboards, ultimately undermining user experience and eroding trust in their integrity.

On the other hand, while prior research on leaderboards has primarily focused on performance-related aspects of ML benchmarks, such as benchmark leakage~\cite{xu2024benchmarking,deng2023benchmark,elangovan2021memorization}, leaderboard plateauing~\cite{maslej2023artificial,ott2022mapping,recht2019imagenet}, or evaluation fairness~\cite{oren2023proving,yang2023rethinking}, the long-term trustworthiness of a leaderboard relies equally on adhering to these key quality attributes mentioned above. This represents a critical gap in the literature: the operational challenges of leaderboard management, particularly in understanding different workflows and identifying recurring pitfalls—termed ``smells''—in leaderboard operations (LBOps), remain underexplored.

Our work aims to uncover patterns and issues undermining trust in FM leaderboards, providing actionable insights to drive improvements, enhance reliability, and promote rigorous model assessment. To achieve these objectives, we seek to answer the following research questions (RQs):
\begin{itemize}
    \item \textbf{RQ$_1$:} \emph{How do FM leaderboards operate?}
    \item \textbf{RQ$_2$}: \emph{What are the issues, or ``smells'', prevalent in the operations of FM leaderboards?} 
\end{itemize}

We employ a three-stage methodology to achieve our research objectives. First, we collect ML leaderboards using the ``leaderboard'' keyword from sources including GitHub, HF Spaces, and PWC. Next, we manually filter out leaderboards that are not relevant to FMs. Finally, we conduct an in-depth examination of each FM leaderboard, analyzing their evaluation processes, documentation, and related publications, and engaging with leaderboard operators where necessary. To ensure comprehensive data analysis, we utilize card sorting~\cite{wood2008card} and negotiated agreement~\cite{campbell2013coding} among the authors, aligning our findings with the RQs.

To our knowledge, this study is the first to explore FM leaderboards as software products, with a focus on their operational lifecycle, termed ``Leaderboard Operations'' (LBOps), and to identify operational issues known as ``leaderboard smells''. We define LBOps as the set of resources and workflows required to rank third-party ML models based on their evaluation performance and to help select the most suitable models for specific contexts. LBOps complements MLOps~\cite{kreuzberger2023machine}, with the former focusing on the submission, evaluation, and comparison of \textbf{third-party} ML models, while the latter addresses the training, versioning, evaluation, and deployment of \textbf{in-house} models.

Based on our analysis of $1,045$ FM leaderboards from five different sources: GitHub, HF Spaces, PWC, spreadsheet and independent platform, our paper makes the following contributions to the SE community:
\begin{itemize}[leftmargin=*]
    \item We derive a domain model for FM leaderboards that highlights the essential components, relationships, and constraints involved in the five different LBOps workflow patterns.
    \item We introduce the novel concept of ``leaderboard smells'', identifying eight unique types of smells that can emerge across nine leaderboard components.
    \item We initiate the \href{https://github.com/SAILResearch/awesome-foundation-model-leaderboards}{``Awesome Foundation Model Leaderboard''} list as a valuable resource for the collection and organization of various FM leaderboards, development toolkits, and publisher organizations. As a side contribution, we also provide the \href{https://huggingface.co/spaces/zhiminy/awesome-foundation-model-leaderboard-search}{``Foundation Model Leaderboard Search''} tool to help users search for specific FM leaderboards.
\end{itemize}
Our works aims to uncover patterns and issues undermining trust in FM leaderboards, providing actionable insights to drive improvements, enhance reliability, and promote rigorous model assessment.
\section{A Software Engineering Perspective on Foundation Model Leaderboards}
\label{sec:usecase}

This section illustrates the SE perspective of FM leaderboards via three persona (Alex, Mia, and Lora) as they navigate the FM leaderboard ecosystem using the fictitious ClearRank leaderboard. It highlights the benefits of transparent and robust leaderboards for operators, model producers, and software engineers, informed by real-world observations and feedback from leaderboard operators during our study.

Alex, the founder of ClearRank, launches a leaderboard to address the widespread frustration among developers over opaque and inconsistent FM rankings, particularly for specialized tasks, such as code completion. In order to deliver clear and actionable evaluations, Alex focuses on building a platform that emphasizes transparency and reliability. In its early stages, ClearRank faces challenges such as automating evaluation workflows and meeting the diverse needs of its users. To address these hurdles, Alex implements robust evaluation protocols and actively engages with the user community to refine the platform. These efforts quickly build user trust, positioning ClearRank as a trusted leader in rigorous FM evaluations.

Mia, a model producer at AIForge, a leading AI company specializing in cutting-edge FMs, submits LogicMaster, her latest fine-tuned FM for reasoning and code completion, to several leaderboards, including ClearRank. Due to varying evaluation criteria, LogicMaster achieves different rankings on different leaderboards. In ClearRank, it excels in reasoning but falls short of multilingual coding benchmarks. Leveraging ClearRank's pairwise evaluation feature, Mia showcases LogicMaster's strengths in direct comparisons with top models. 
  
Lora, a software engineer at InnovateTech, faces the challenge of selecting the best FM for a new code completion tool. As a mid-sized tech company, InnovateTech lacks the resources for manual FM evaluation and struggles with inconsistencies across leaderboards. Drawn to ClearRank's transparent evaluations, Lora identifies LogicMaster's strong performance in reasoning tasks, which aligns with the critical requirements of her project. To ensure the model's suitability for real-world scenarios, she leverages ClearRank's pairwise evaluations to directly compare LogicMaster against competing models on task-specific examples. The detailed insights from these comparisons provide the confidence she needs, leading her to select LogicMaster as a reliable solution for InnovateTech.

This example highlights the importance of establishing clear workflows and best practices in leaderboard operations while addressing common issues, such as opaque evaluation protocols, to enhance reliability and trust within the ML community. It also prompts broader questions:
\begin{enumerate}
    \item What are the common operational workflows on leaderboards and what are their strengths and weaknesses?
    \item What components and practices are essential for defining and maintaining these workflows?
    \item What operational ``smells'' compromise the reliability and trustworthiness of the leaderboards?
\end{enumerate}
Answering these questions is vital for enhancing the FM leaderboard ecosystem, promoting transparency, reliability, and usability for all stakeholders.
\section{Background and Related Work}
\label{sec:background}

While most of the SE research involving FMs has focused on leveraging FMs to enhance SE tasks, such as code completion~\cite{dakhel2023github,barke2023grounded}, code understanding~\cite{nam2024using}, program repair~\cite{fan2023automated}, and software development~\cite{qian2023communicative}, or utilizing SE techniques to refine the FM development process, such as ChainForge~\cite{arawjo2023chainforge}, AI2Apps~\cite{pang2024ai2apps}, and SPADE~\cite{shankar2024spade}, our study presents a unique perspective on FM leaderboards within the SE context. Positioned in the SE4AI domain~\cite{mcdermott2020ai4se}, our research critically examines the workflows and practices that FMs undergo to appear on leaderboards. These systems are increasingly influential, but face significant challenges in ensuring fairness, reliability, and accountability. Recently, the $2024$ AI Index Report~\cite{maslej2024artificialintelligenceindexreport} underscores the lack of standardization in FM evaluations, preventing fair comparison of the best models across benchmarks. Based on these concerns, our study proposes a systematic approach to identify key issues in LBOps. Our research promotes responsible FM comparison and advocates for accountability among leaderboard stakeholders—designers, architects, developers, testers, maintainers, managers, and publishers, collectively termed ``leaderboard operators''—to enhance the reliability and long-term utility of FM leaderboards.

This exploration of leaderboards connects to broader discussions on ranking systems across domains, including behavioral psychology, human-computer interaction, and AI. For example, Höllig~et~al.~\cite{hollig2020individualizing} examine the influence of trait competitiveness – an individual's inherent inclination to engage in competitive activities – and leaderboard design on individual performance and engagement within gamified systems, highlighting the importance of these factors in shaping user experiences. Furthermore, Na~et~al.~\cite{na2023leaderboard} explore how leaderboard positions affect competence satisfaction, which, in turn, affects motivation levels and task persistence. Kabongo~et~al.~\cite{kabongo2021automated} highlight the difficulties in tracking scientific progress in the AI community due to the large volume of research publications. In response, specialized software has been developed, including Axcell~\cite{kardas2020axcell}, TELIN~\cite{yang2022telin}, and ORKG~\cite{kabongo2023orkg} to automatically extract leaderboard data from publications, thus reducing the dependency on labor-intensive human annotation. Furthermore, Singh~et~al.~\cite{singh2024legobench} address the challenge of information overload in scientific research by providing a benchmark to evaluate these systems that generate scientific leaderboards. 

However, despite such efforts, several performance concerns in ML evaluations remain unresolved. These include benchmark leakage~\cite{xu2024benchmarking,deng2023benchmark,elangovan2021memorization}, which compromises the integrity of test results; leaderboard plateauing~\cite{maslej2023artificial,ott2022mapping,recht2019imagenet}, where progress stagnates due to saturated benchmarks; and evaluation fairness~\cite{oren2023proving,yang2023rethinking}, which questions the consistency and equity of scoring mechanisms. To address these challenges, Chiang~et~al.~\cite{chiang2024chatbot} propose the Chatbot Arena which uses pairwise comparison methods to enhance the reliability and fairness of FM evaluations. While such approaches provide promising alternatives to traditional leaderboard rankings, our study takes a different perspective. We focus on analyzing LBOps by examining their workflow patterns, developing a domain model, and identifying recurring issues with the goal of promoting standardization and fostering responsible FM comparisons.
\section{Methodology}
\label{sec:method}

\subsection{Research Questions}
\label{sec:method:rq}

Our research aims to improve the sustainability and trustworthiness of FM leaderboards by addressing two RQs.
\begin{itemize}[leftmargin=*]
    \item \textbf{RQ$_1$:} \emph{How do FM leaderboards operate?} 
    To enhance the effectiveness of FM leaderboards, it is essential to gain insights into their operational aspects. Thus, this RQ investigates the workflow patterns and domain concepts necessary to maintain the functionality and usefulness of leaderboards. To address this RQ, we analyze FM leaderboards' submission/contribution protocols, documentation/publication (\ie, blogs, reports), and commit history to identify patterns of leaderboard operations (LBOps). In parallel, we derive the domain model for key concepts in LBOps, capturing its major components and their relationships. This model helps stakeholders gain a clearer understanding of LBOps' structure, identify optimization opportunities, and support its evolution to address emerging requirements and adapt to diverse environments.
    \item \textbf{RQ$_2$}: \emph{What are the issues, or ``smells'', prevalent in the operations of FM leaderboards?} 
    Leveraging our understanding of LBOps, this RQ aims to identify and categorize common operational issues within FM leaderboards, termed as ``leaderboard smells''. Inspired by the concepts of ``code smells''~\cite{sharma2018survey}, ``design smells''~\cite{suryanarayana2014refactoring}, and ``architectural smells''~\cite{garcia2009identifying}, a leaderboard smell is an operational issue that hampers the leaderboard's functionality or sustainability, often leading to dissatisfaction among users. By examining the characteristics and distribution of these smells across different sources and workflow patterns, we aim to provide actionable insights that enable leaderboard operators to anticipate and mitigate similar pitfalls in future development.
\end{itemize}

\subsection{Study Design}
\label{sec:method:design}

\begin{figure*}
\centering
\includegraphics[width=\linewidth]{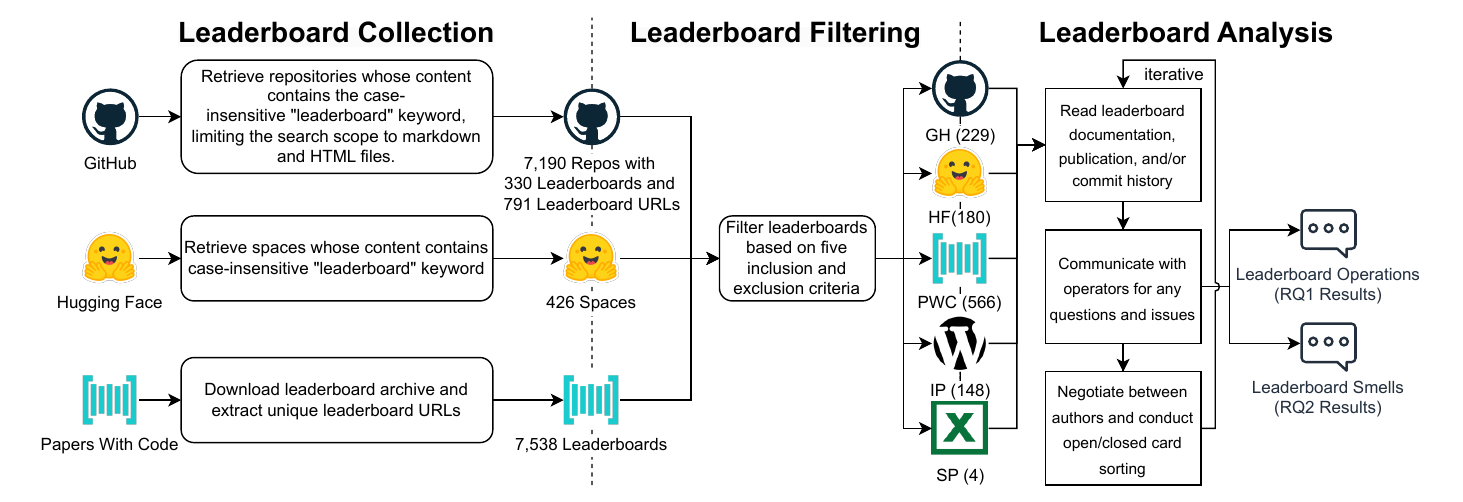}
\caption{Three-phase study workflow: (1) Leaderboard Collection – Collect ML leaderboards from GitHub, HF Spaces, and PWC, to build a comprehensive dataset; (2) Leaderboard Filtering – Apply predefined inclusion/exclusion criteria to manually review and curate the collected leaderboards; (3) Leaderboard Analysis – Investigate leaderboard documentation, evaluation methodologies, and operational workflows, engaging with operators to derive actionable insights.}
\label{fig:study workflow}
\centering
\end{figure*}

Figure~\ref{fig:study workflow} outlines our three-phase study workflow.

\subsubsection{Phase 1: Leaderboard Collection}

We target three primary platforms for FM leaderboards: \href{https://github.com}{GitHub}, \href{https://huggingface.co/spaces}{Hugging Face (HF) Spaces}, and \href{https://paperswithcode.com/sota}{Papers With Code (PWC)}. For GitHub, we use the \href{https://sourcegraph.com/code-search}{SourceGraph Code Search API} to retrieve repositories containing the case-insensitive ``leaderboard'' keyword in their content. Based on our observations, ML leaderboards or links redirecting to them are typically found in markdown (\textbf{.md}) files or GitHub Pages (\textbf{.html}) hosted within GitHub repositories. To optimize leaderboard retrieval, we apply these two file extension constraints, which results in $7,190$ repositories. The first two authors randomly select $720$ ($\sim10\%$) repositories for manual inspection to identify URLs that directly link to ML leaderboards or the websites hosting them. Initially, three cases of disagreement arise regarding what qualifies as an ML leaderboard, but these are resolved after another discussion between the authors. Afterwards, the first author independently reviews the remaining repositories to identify any URLs redirecting to ML leaderboards. 

For GitHub, we identify $1,681$ mentions of the ML leaderboards, with $1,121$ unique entries. Among these, $330$ leaderboards are hosted directly on scraped GitHub repositories, while $791$ URLs redirect to leaderboards hosted elsewhere. For HF Spaces, we retrieve $429$ spaces containing the case-insensitive ``leaderboard'' keyword. For PWC, we download the leaderboard archive directly from the \href{https://production-media.paperswithcode.com/about/evaluation-tables.json.gz}{official portal}, retrieving a total of $7,539$ leaderboards. Occasionally, we discover new FM leaderboards within the identified leaderboard documentation. For example, \href{https://redarena.ai/leaderboard}{RedTeam Arena} is recommended on the front page of Chatbot Arena. Using the backward snowball approach~\cite{jalali2012systematic}, we identify $7$ additional FM leaderboards.

\subsubsection{Phase 2: Leaderboard Filtering}

Subsequently, we implement a systematic process to refine and filter the collected leaderboard-related resources. The first two authors start by conducting a random check of $110$ leaderboard mentions, $40$ HF spaces, and $750$ PWC leaderboards ($\sim10\%$ of the total) to evaluate their compliance with the inclusion and exclusion criteria (discussed below). This cross-check identifies four disagreements, which are promptly resolved through negotiated agreement. Once the criteria are finalized, the first author systematically applies them to the remaining dataset:
\begin{itemize}[leftmargin=*]
    \item \emph{Exclusion of Non-leaderboards}: This criterion ensures that our study focuses exclusively on ML leaderboards, adhering to the definition provided by Hugging Face. For instance, some spaces (\eg, \href{https://huggingface.co/spaces/leaderboards/LeaderboardsExplorer}{Leaderboard Explorer}) on HF include ``leaderboard'' in their titles but are actually tools for developing ML leaderboards. Overall, we exclude $50$ spaces from HF Spaces: 6 leaderboard development kits, $29$ leaderboard templates, and $15$ empty spaces, resulting in $379$ spaces classified as leaderboards.
    \item \emph{Inclusion of Leaderboards with FM Evaluations}: This criterion ensures that our study focuses exclusively on ML leaderboards that evaluate FMs (or FM-powered agents). In our study, we define FMs as ML models with at least one billion parameters, following the widely accepted definition and standard established by the AI community~\cite{bommasani2024foundation} and USA government~\cite{biden2023executive}. To identify FM evaluations, we manually inspect the available evaluation records for columnar attributes, including model name, parameter count, and other provenance information. In total, we discover $229$ FM leaderboards from GitHub, while excluding $6,969$ from PWC and $7$ from HF Spaces.
    \item \emph{Exclusion of Duplicate Leaderboards}: This criterion ensures that our analysis avoids unnecessary redundancies. For instance, several leaderboards have been forked from the Chatbot Arena leaderboard with minimal or no modifications\footnote{\url{https://huggingface.co/spaces?search=chatbot+arena+leaderboard}}. However, if these forked leaderboards introduce new evaluations that differ from the original, we still include them in our analysis. Applying this criterion, we have excluded $94$ duplicate spaces from HF Spaces.
    \item \emph{Exclusion of Fully Malfunctioning Leaderboards}: This criterion ensures that our analysis excludes leaderboards with persistent runtime errors, prolonged unresponsiveness, operator pauses, or no evaluation records. For example, the ``image-generation-on-celeba-3'' leaderboard remained perpetually loading throughout our study\footnote{\url{https://github.com/paperswithcode/sota-extractor/issues/39}}. However, if a leaderboard is fully malfunctioning on one platform but remains functional on another, it is retained for further analysis. Overall, we exclude one unresponsive leaderboard and 80 leaderboards without any evaluations from PWC, along with $101$ HF spaces exhibiting persistent runtime errors, verified through bi-weekly rechecks.
    \item \emph{Exclusion of Unlaunched Leaderboards}: This criterion ensures that our analysis remains focused on launched leaderboards. For example, at the time of our study, the \href{https://huggingface.co/spaces/cambioml/parser-leaderboard/discussions/2}{Parser Arena leaderboard} is still under construction. However, if only a subset of evaluations is incomplete on a leaderboard, we still include it in our analysis as long as the overall leaderboard remains functional. In total, we exclude $4$ incomplete leaderboards from HF Spaces.
\end{itemize}

We notice that multiple leaderboards are sometimes grouped under the umbrella of a higher-level leaderboard. For example, the \href{https://rank.opencompass.org.cn}{Large Language Model Leaderboard} contains the ``CompassBench Leaderboard'', ``CompassAcademic Leaderboard'', and ``Compass Arena Leaderboard''. In such cases, we count the former as a single entity, rather than treating its descendants as separate entries. However, if multiple leaderboards on the same website lack a unified name at the highest level, we treat them as separate leaderboards. For example, the \href{https://www.superclueai.com}{SuperCLUE series of leaderboards} are hosted on the same website but do not have a unified name for their individual leaderboards.

\begin{figure}
\centering
\includegraphics[width=\linewidth]{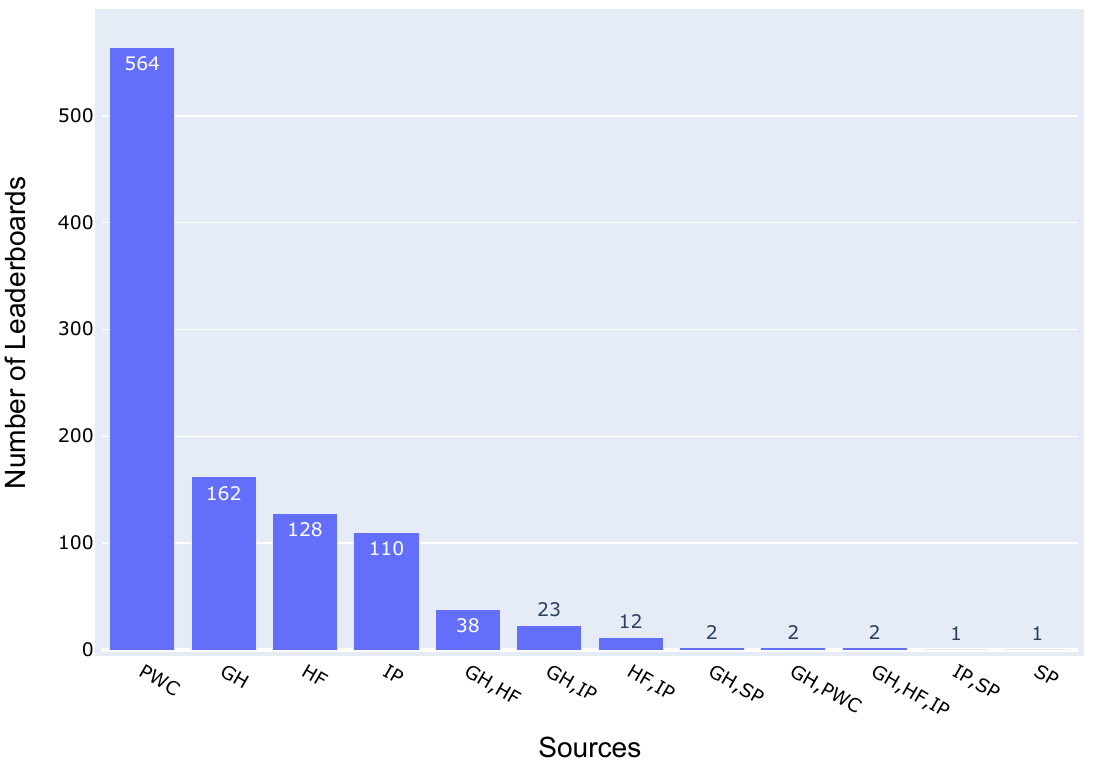}
\caption{Distribution of FM leaderboards across various different sources. The abbreviations used are: GH (GitHub), HF (Hugging Face Spaces), PWC (Papers With Code), IP (independent platform), and SP (spreadsheet platform). Comma-separated names indicate leaderboards hosted on multiple sources.}
\label{fig:sources}
\centering
\end{figure}

After this phase, our approach has identified $1,045$ unique FM leaderboards. Figure~\ref{fig:sources} illustrates the distribution of these leaderboards across different sources. In particular, PWC is the most popular platform, hosting $54.16\%$ ($566/1,045$) of the leaderboards, followed by GitHub ($21.82\%$), HF ($17.22\%$), and independent platforms—websites hosted by third-party organizations—at $14.16\%$. Spreadsheet-based leaderboards constitute the remaining $0.38\%$. We find that $7.66\%$ ($80/1,045$) FM leaderboards are hosted across multiple sources. Among these, GitHub and HF Spaces stand out as the most common pairing, accounting for $47.50\%$ ($38/80$).

To enhance outreach, we share our leaderboard collection through the \href{https://github.com/SAILResearch/awesome-foundation-model-leaderboards}{Awesome Foundation Model Leaderboards list}. Additionally, we provide a \href{https://huggingface.co/spaces/zhiminy/awesome-foundation-model-leaderboard-search}{search tool} to help stakeholders efficiently discover leaderboards aligned with their interests. For a comprehensive compilation of the identified FM leaderboards, complete with metadata, we direct readers to our online replication package~\cite{replication_package}.

\subsubsection{Phase 3: Leaderboard Analysis}

\emph{$RQ_1$}: We start by reviewing each collected leaderboard's documentation and publications (\eg, blogs, reports) and, where available, its commit history to identify typical activities and their stakeholders, such as those who submit model artifacts, model outputs, or evaluation records for integration into new or existing leaderboards. Specifically, we outline stakeholder roles, input artifacts, and generated outputs in each identified activity. When leaderboard information is unclear, we reach out to its operators through designated discussion platforms, including email, social networks (\eg, \href{https://discord.com}{Discord}, \href{https://slack.com}{Slack}, and \href{https://web.wechat.com}{WeChat}, and discussion forums (\eg, GitHub issues, HF Spaces discussions), often requiring multiple rounds of communication. By \collectdate, we have initiated $834$ discussions on GitHub, with $422$ of those receiving responses, and $651$ discussions on HF Spaces, with $263$ replies. We also sent $14$ emails, mainly to PWC operators, receiving $8$ responses. Furthermore, we conducted around $50$ rounds of conversations on WeChat and $4$ rounds of discussion on Discord and Slack. 

Table~\ref{tab:leaderboard statistics} summarizes key findings from our leaderboard exploration. We find that only $35.12\%$ ($367/1045$) of FM leaderboards are linked to specific GitHub repositories, indicating a significant gap in transparency and traceability for leaderboard implementations. In particular, all spreadsheet-based leaderboards ($100\%$) have associated GitHub repositories. In contrast, $99.65\%$ PWC leaderboards lack an associated GitHub repository. Furthermore, we observed that $99.14\%$ of the leaderboards feature only one major release to date, where a major release signifies significant changes, improvements, or initial creation. This reflects either limited active maintenance of the leaderboard or indicates that this domain is still in its early stages. For PWC leaderboards, it is generally not possible to determine the presence of major releases, as most (with only two exceptions) do not provide any information about maintainers or version history. Additionally, $76.65\%$ ($801/1045$) FM leaderboards include explicit submission channels or protocols for model-related artifacts, such as model API portals, prediction files, or evaluation records. These channels are essential for standardized and seamless submissions that ensure the integrity and comparability of results. Notably, all PWC leaderboards ($100\%$) provide submission channels where registered users can submit, edit, or remove evaluations directly, whereas only $25\%$ ($1/4$) of spreadsheet-based leaderboards allow user submissions. Furthermore, only $1.44\%$ ($15/1045$) FM leaderboards support submissions beyond model-related artifacts, such as benchmarks or evaluators. This highlights operators' efforts to address challenges, such as ``leaderboard plateauing''~\cite{ott2022mapping,maslej2023artificial} and ``inefficient benchmarking''~\cite{polo2024tinybenchmarks}, aiming to enhance the adaptability and continuous evolution of leaderboards. However, neither PWC nor spreadsheet-based leaderboards support such submissions. Lastly, we found that $18.66\%$ ($195/1,045$) of FM leaderboards lack links to relevant publications, codebases, or websites for the evaluated models. In contrast, PWC leaderboards consistently provide model provenance information by default. This lack of transparency and traceability raises concerns about the trustworthiness and reliability of these evaluation records, underscoring the urgent need for improved documentation practices in LBOps.

\begin{table}[t]
\centering
\caption{Detailed statistics of key attributes of FM leaderboards across different sources, with ``NA'' indicating attributes not applicable to specific leaderboards.}
\label{tab:leaderboard statistics}
\begin{NiceTabular}{l|C{1cm}C{1cm}C{1cm}C{1cm}C{1cm}}
& \multicolumn{5}{c}{\textbf{Leaderboard Statistics}} \\ 
\cline{2-6}
\textbf{Source} & \rotatebox{90}{\makecell{GitHub Repository \\ Available?}} & \rotatebox{90}{\makecell{Major Releases \\ Available?}} & \rotatebox{90}{\makecell{Submission Channel/Protocol \\ Available?}} & \rotatebox{90}{\makecell{Allows Other \\ Submission Types?}} & \rotatebox{90}{\makecell{Model Provenance \\ Available?}} \\ \midrule
GitHub & 100.00\% (229/229) & 1.31\% (3/229) & 44.98\% (103/229) & 2.18\% (5/229) & 51.97\% (119/229) \\ \midrule
HF Spaces & 67.78\% (122/180) & 3.33\% (6/180) & 55.00\% (99/180) & 4.44\% (8/180) & 66.67\% (120/180) \\ \midrule
\makecell{independent \\ platform} & 59.46\% (88/148) & 1.35\% (2/148) & 52.03\% (77/148) & 4.73\% (7/148) & 62.84\% (93/148) \\ \midrule
PWC & 0.35\% (2/566) & NA & 100.00\% (566/566) & 0.00\% (0/566) & 100.00\% (566/566) \\ \midrule
\makecell{spreadsheet \\ platform} & 100.00\% (4/4) & 0.00\% (0/4) & 25.00\% (1/4) & 0.00\% (0/4) & 75.00\% (3/4) \\ \midrule
Overall & 35.12\% (367/1045) & 1.88\% (9/479) & 76.65\% (801/1045) & 1.44\% (15/1045) & 81.34\% (850/1045) \\
\end{NiceTabular}
\end{table}

We also hold weekly author meetings to review findings, refine operators' insights, and identify recurring patterns. These discussions foster collaborative consensus on the key components of LBOps workflows. Through this iterative process, we ultimately identify five workflow patterns applicable to all FM leaderboards. Notably, PWC leaderboards primarily follow a standardized default workflow pattern (\ie, $P_1$), as noted in the \href{https://github.com/paperswithcode/axcell}{PWC scraping tool documentation}. During the collection phase, only two exceptions for PWC leaderboards are identified through analysis of the GitHub repository. Consequently, our closed card sorting analysis is limited to non-PWC leaderboards. To ensure a representative analysis~\cite{mokkink2023sample,liddy2011methods}, the first two authors randomly select $100$ leaderboard samples (representing $20.79\%$ of the total) and independently assign composite labels based on the five identified workflow patterns. This process yields a Cohen's kappa inter-rater reliability score of $0.963$, with only two discrepancies, indicating a very high level of agreement between the raters. After resolving any disagreements, the first author continues assigning pattern labels to the remaining non-PWC leaderboards. 

While identifying the workflow patterns, we simultaneously develop a domain model that encapsulates the core concepts involved in the LBOps workflows. The model organizes key entities, their attributes, and their relationships, specifying interactions such as ``users submit models'', ``users upload predictions or evaluation results'', and ``leaderboards integrate and rank models based on evaluations''. Workflow actions such as ``submit'', ``evaluate'', and ``integrate'' are explicitly mapped to the domain model, along with their corresponding input (\eg, model files, prediction outputs, evaluation metrics) and output artifacts (\eg, evaluation records, ranking dataframes). To enhance clarity, we document the domain model in a class diagram to illustrate these elements and their interactions comprehensively. Lastly, we validate our domain model with the lead operator of \href{https://www.superclueai.com}{SuperCLUE}, one of China's most prominent leaderboards, leveraging his expertise in long-term leaderboard maintenance. His insightful suggestion to break the domain model into layers enhances its alignment with the identified workflow patterns, ensuring both accuracy and comprehensiveness.

\emph{$RQ_2$}: In our study, we define a ``leaderboard smell'' as a recurring operational issue that undermines key non-functional requirements of a leaderboard~\cite{offutt2002quality,barbacci1995quality}, such as reliability, availability, maintainability, and usability. This definition draws inspiration from software engineering ``smells'', which serve as indicators of technical debt or systemic issues~\cite{sharma2018survey,suryanarayana2014refactoring,garcia2009identifying,kim2020empirical,rahman2019seven,wu2023systematic,jafari2021dependency}. Specifically, we classify leaderboard behaviors as smells only when they reflect persistent or recurring problems, rather than temporary issues, such as \href{https://github.com/goML-offers/doc_attribute/issues/7}{brief website outages}.

As noted in $RQ_1$, we submit issue reports during the leaderboard filtering and workflow investigation process through various communication channels, including GitHub, HF Spaces, email, and social media platforms. Based on feedback from leaderboard operators, validation from other users, and weekly iterative discussions among the authors, we identify and confirm $476$ of these reports as smell cases: $257$ from GitHub and $217$ from HF Spaces. Of the identified smell cases, $43.49\%$ ($207/476)$ have been resolved through operator interventions or our proposed fixes, including pull requests. Specifically, we contributed $13$ pull requests on GitHub (10 accepted) and $7$ on HF Spaces ($4$ accepted). Additionally, we have directly updated $45$ PWC leaderboards to address our identified smells. Among the unresolved cases, $65$ ($24.16\%$) are acknowledged and confirmed by operators, while $3$ ($1.12\%$) were independently validated by other users. Notably, none of the identified smell cases have been refuted by the operators to date. We also notice that $33$ ($6.93\%$) smell cases are explicitly acknowledged by operators as technical debt~\cite{potdar2014exploratory} (SATD), namely, the issues are known but left unaddressed on purpose. An example is the HELM Classic leaderboard, where two identical ranking dataframes appear (redundant entity smell) due to \href{https://github.com/stanford-crfm/helm/issues/2351}{a computational issue by the operators}.

Our discussions with practitioners provide nuanced insights into the types and resolutions of leaderboard smells, enabling us to cluster cases into categories based on their prominent features and similarities. To mitigate bias, we apply the ``negotiated agreement'' method, a technique commonly used in empirical software engineering~\cite{winter2022developers,chattopadhyay2019latent,johnson2022empirical,fazzini2022characterizing,da2020dominoes,chattopadhyay2020tale}. This method involves multiple researchers independently reviewing the data, identifying discrepancies in their analyses or categorizations, then collaboratively resolving these via discussion to reach a mutually agreed-upon conclusion~\cite{campbell2013coding}. For ambiguous smell cases, we seek further clarification from leaderboard operators to refine our categorization. 

This process, spanning from January to June 2024 and September to October 2024, involves weekly discussions among the authors. Through this iterative approach, we finally identify eight unique types of smells that account for $92.02\%$ ($438/476$) of our issue reports. The remaining are categorized as ``others'', as they represent more traditional software smells, such as \href{https://huggingface.co/spaces/mii-llm/open_ita_llm_leaderboard/discussions/7}{typographic} or \href{https://github.com/THU-KEG/KoLA/issues/9}{authentication} errors. The complete set of cases, along with their associated smells, is available in our replication package~\cite{replication_package}. In this package, URLs are color-coded for clarity: red for resolved cases, green for unresolved but confirmed by operators, yellow for unresolved but user-confirmed, and bold for SATDs.

Similarly to $RQ_1$, we obtain the feedback from the lead operator of \href{https://www.superclueai.com}{SuperCLUE} in our catalog of leaderboard smells. His review affirms the relevance and credibility of our findings, while emphasizing the need for efficient, sustainable leaderboard management practices to meet long-term maintenance and community expectations.
\section{$RQ_1$ Results: Leaderboard Operations}
\label{sec:rq1-results}

This section presents our identified workflow patterns in LBOps, as well as the corresponding domain model that we develop in parallel.

\subsection{Workflow Patterns}

\begin{figure}
\centering
\includegraphics[width=\linewidth]{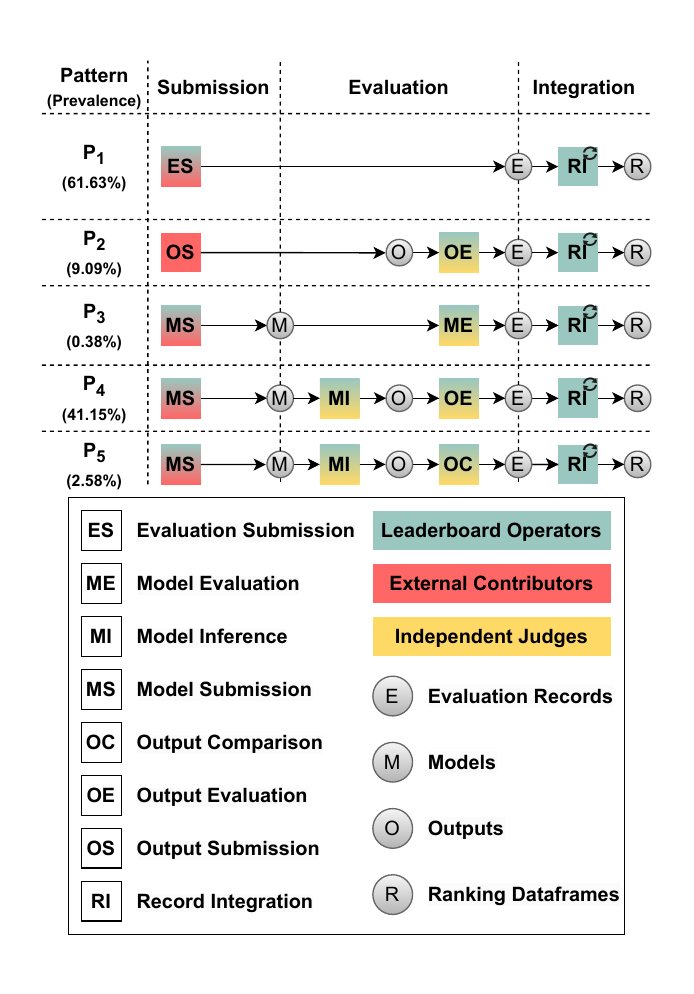}
\caption{Schematic representation of workflow patterns in LBOps, ordered by the number of operations. The arrow indicates the execution sequence; the block represents an operation; the circle denotes an artifact or access to it; the color signifies a role (mixed colors indicate multiple possible roles); the loop symbol marks a continuous integration process.}
\label{fig:workflow pattern}
\centering
\end{figure}

Figure~\ref{fig:workflow pattern} presents the five identified workflow patterns and their respective prevalence within the FM leaderboards. Each of them spans three major phases: artifact \textit{submission}, model \textit{evaluation}, and record \textit{integration}. In the following sections, we provide an in-depth explanation of each workflow pattern, leaderboard examples, and a discussion of their characteristics.

$P_{\useworkflowIndex}$ \textbf{External Evaluation Integration} \\ 
\emph{Rationale}: Leaderboard operators (\raisebox{-0.5mm}{\adjustbox{height=0.8\baselineskip,valign=m}{\includegraphics{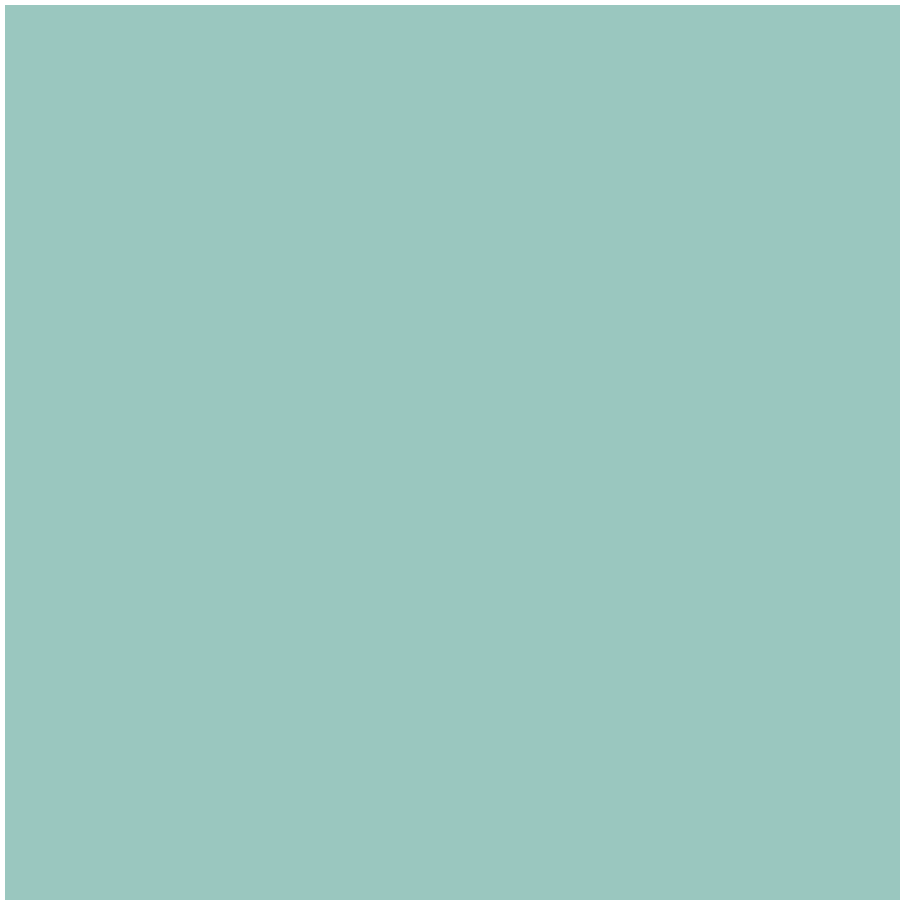}}}) and/or external contributors (\raisebox{-0.5mm}{\adjustbox{height=0.8\baselineskip,valign=m}{\includegraphics{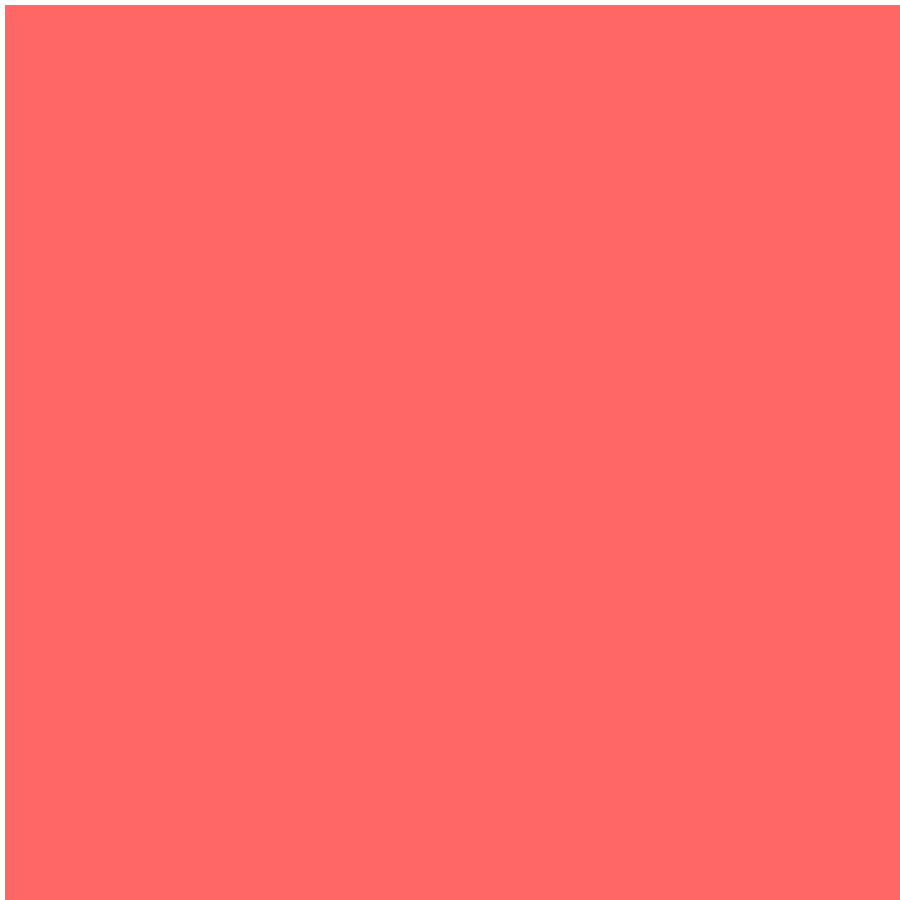}}}) collect evaluation records (\raisebox{-0.5mm}{\adjustbox{height=0.8\baselineskip,valign=m}{\includegraphics{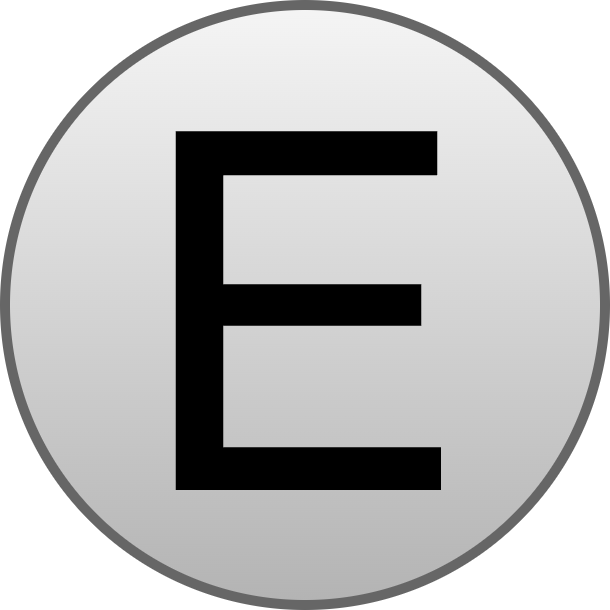}}}) from online sources, such as research articles and model cards. Alternatively, external contributors can independently evaluate their models, generating the evaluation records according to the evaluation steps outlined on the leaderboard website. Then they submit these evaluation records (\raisebox{-0.5mm}{\adjustbox{height=0.8\baselineskip,valign=m}{\includegraphics{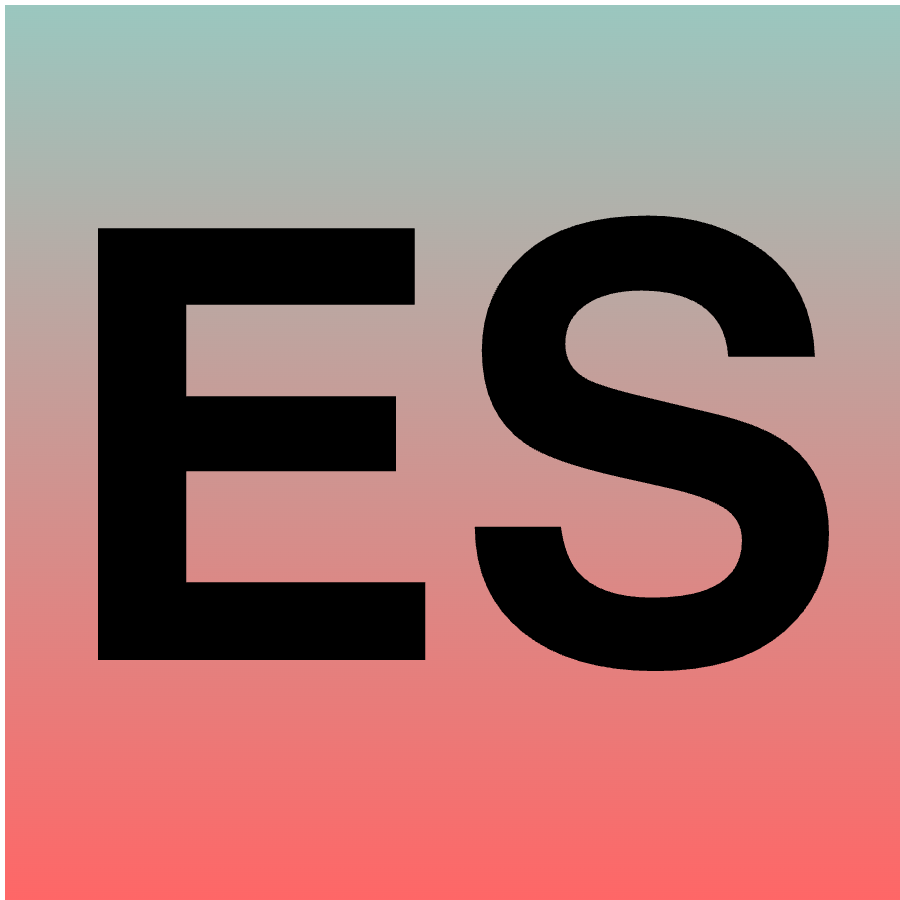}}}) through designated channels, including emails, issue reports, pull requests, and submission portals, to the leaderboard. Afterwards, the leaderboard operators can optionally review the submissions and integrate (\raisebox{-0.5mm}{\adjustbox{height=0.8\baselineskip,valign=m}{\includegraphics{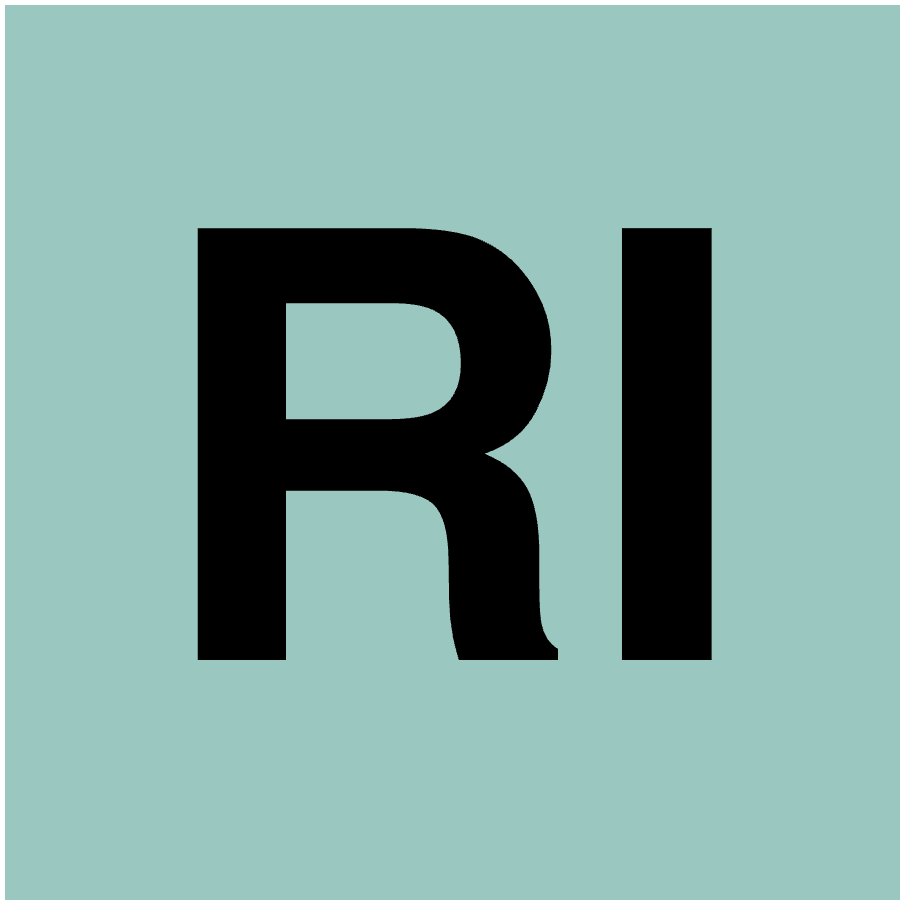}}}) them into new or existing ranking dataframes (\raisebox{-0.5mm}{\adjustbox{height=0.8\baselineskip,valign=m}{\includegraphics{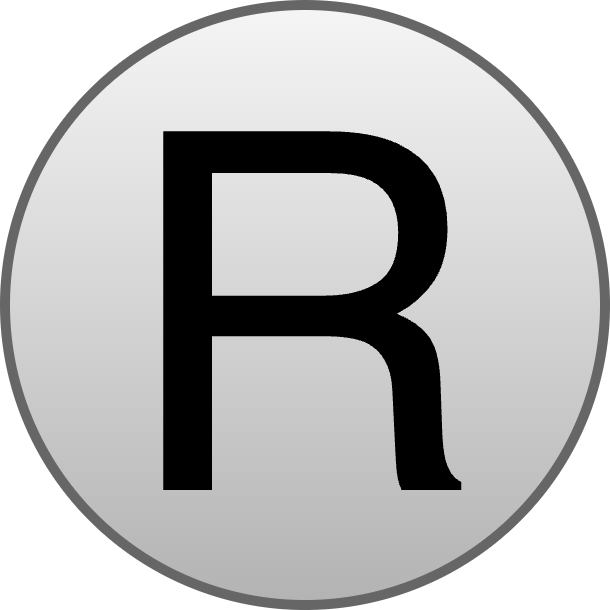}}}), whose definition is detailed in Section~\ref{sec:rq1-results:domain-model}. In some cases, external contributors can directly integrate their evaluation records without requiring further approval. The continuous nature of the record integration process (\raisebox{-0.5mm}{\adjustbox{height=0.8\baselineskip,valign=m}{\includegraphics{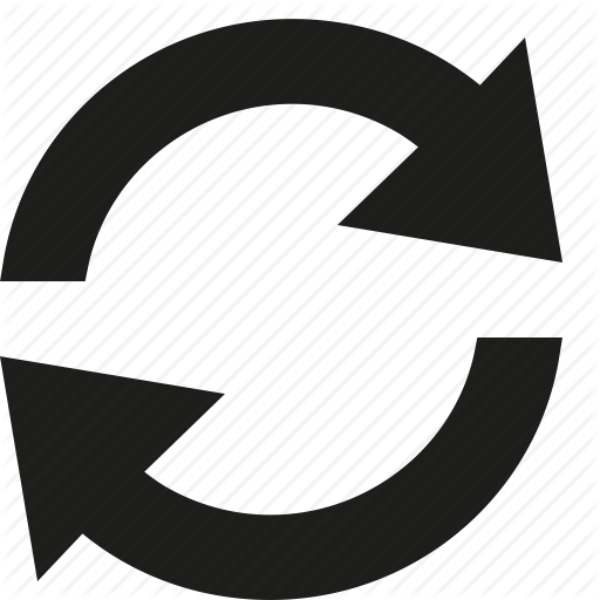}}}) ensures leaderboards to remain current and relevant. \\
\emph{Example}: \href{https://tatsu-lab.github.io/alpaca_eval}{AlpacaEval}, \href{https://huggingface.co/spaces/bigcode/bigcode-models-leaderboard}{Big Code Models Leaderboard}, \href{https://github.com/LudwigStumpp/llm-leaderboard}{LLM-Leaderboard}. \\
\emph{Discussion}: From a computational perspective, evaluation integration is the least taxing workflow for leaderboard operators, as their role is primarily limited to verifying submitted evaluations rather than conducting the evaluations themselves. However, the accuracy and reliability of the submitted evaluations depend on the credibility of their sources. Without a rigorous review process, this reliance on external submissions may undermine the trustworthiness of the rankings. To address these quality concerns, some leaderboards, such as \href{https://mathvista.github.io/\#leaderboard}{MathVista}, require external contributors to submit both score and output files from FM evaluations, allowing operators to independently verify submission. In contrast, leaderboards with minimal review mechanisms, such as PWC, allow registered users to freely submit, modify, or remove evaluations. This flexibility can introduce common issues or ``smells'', as discussed in Section~\ref{sec:rq2-results}.

$P_{\useworkflowIndex}$ \textbf{Model Output Evaluation} \\ 
\emph{Rationale}: External contributors (\raisebox{-0.5mm}{\adjustbox{height=0.8\baselineskip,valign=m}{\includegraphics{figures/rq1/EC.png}}}) run the benchmark test set on their models and then submit (\raisebox{-0.5mm}{\adjustbox{height=0.8\baselineskip,valign=m}{\includegraphics{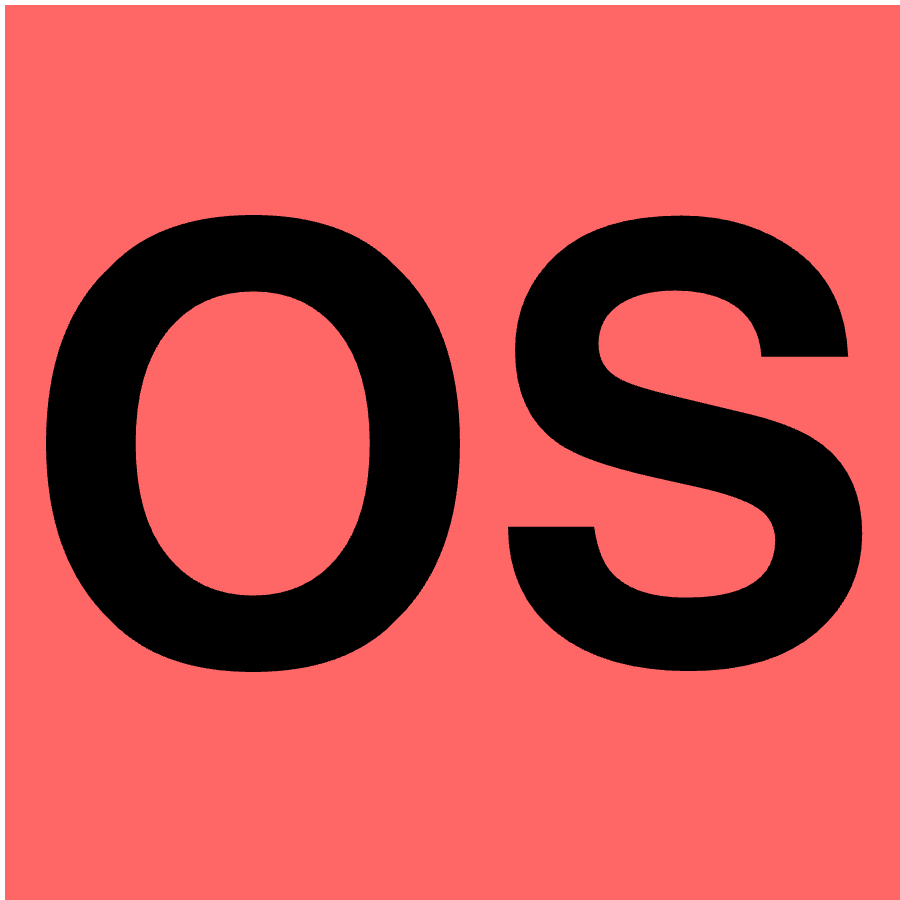}}}) the resulting output (\raisebox{-0.5mm}{\adjustbox{height=0.8\baselineskip,valign=m}{\includegraphics{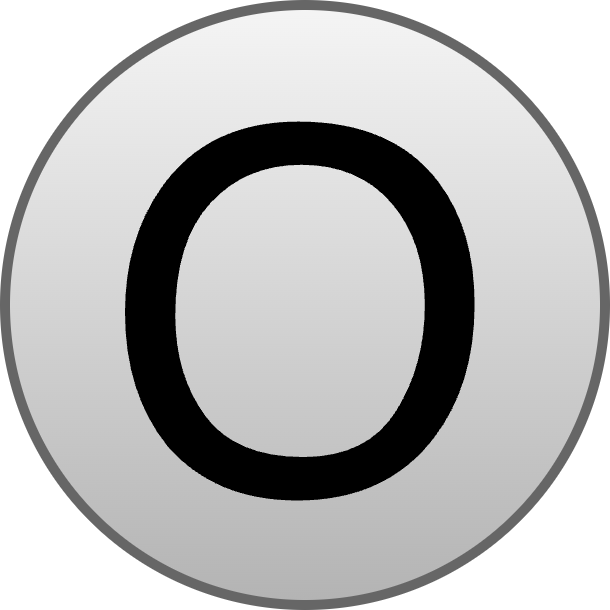}}}) through designated channels to the leaderboard. These outputs are subsequently evaluated (\raisebox{-0.5mm}{\adjustbox{height=0.8\baselineskip,valign=m}{\includegraphics{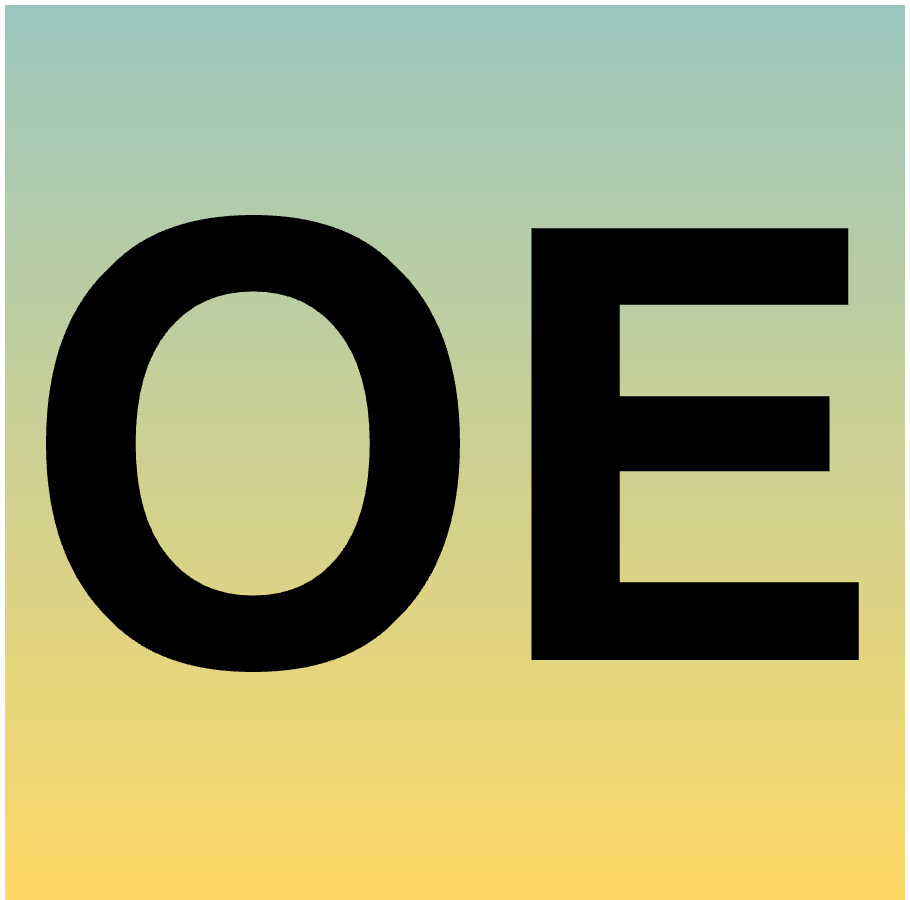}}}) by leaderboard operators (\raisebox{-0.5mm}{\adjustbox{height=0.8\baselineskip,valign=m}{\includegraphics{figures/rq1/LO.png}}}) against benchmark ground truths. If ground truths are not available, independent judges (\raisebox{-0.5mm}{\adjustbox{height=0.8\baselineskip,valign=m}{\includegraphics{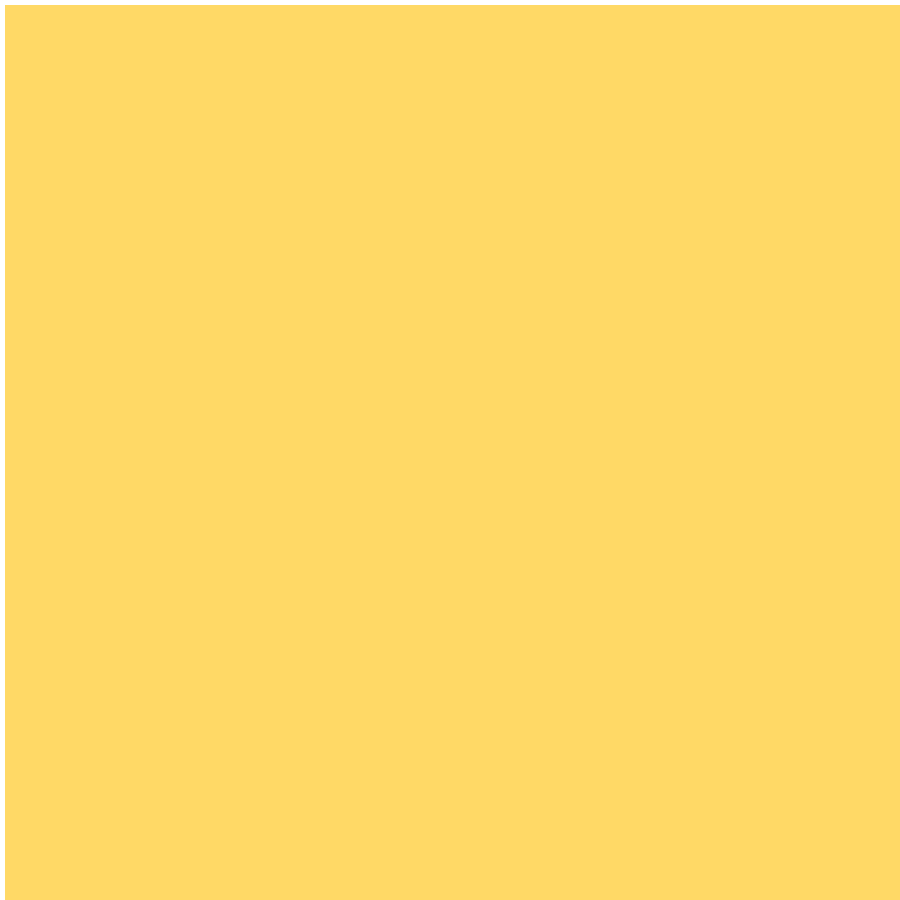}}})—either AIs (\eg, AlignBench~\cite{liu2023alignbench}) or humans (\eg, Human-as-a-Judge~\cite{zhuge2024agent})—act as reviewers to directly assign scores to the models following predefined benchmark protocols. Finally, the evaluation scores as records (\raisebox{-0.5mm}{\adjustbox{height=0.8\baselineskip,valign=m}{\includegraphics{figures/rq1/E.png}}}) are integrated (\raisebox{-0.5mm}{\adjustbox{height=0.8\baselineskip,valign=m}{\includegraphics{figures/rq1/RI.png}}}) into new or existing ranking dataframes (\raisebox{-0.5mm}{\adjustbox{height=0.8\baselineskip,valign=m}{\includegraphics{figures/rq1/R.png}}}) for future releases. The continuous nature of record integration process (\raisebox{-0.5mm}{\adjustbox{height=0.8\baselineskip,valign=m}{\includegraphics{figures/rq1/CI.png}}}) ensures leaderboards to remain current and relevant. \\
\emph{Example}: \href{https://leaderboard.allenai.org}{AI2 leaderboards}, \href{https://huggingface.co/spaces/gaia-benchmark/leaderboard}{GAIA}, \href{https://wilds.stanford.edu/leaderboard}{WILDS}. \\
\emph{Discussion}: Output evaluation improves scalability for leaderboard operators by shifting the computational burden of running benchmark tests to external contributors, allowing the leaderboard to accommodate a larger number of submissions. However, this scalability comes at the cost of higher entry barriers for individuals or teams, as they must independently execute benchmark tests and adhere to submission protocols. For novice contributors, complex submission processes could reduce leaderboard usability, highlighting the importance of user-friendly interfaces and comprehensive documentation to lower the barrier to participation. Furthermore, this workflow inherently carries a risk of manipulation, as contributors may intentionally fine-tune evaluation settings or model outputs to artificially boost their rankings. This might compromise the integrity of the leaderboard and undermine fair competition by shifting the focus away from genuine model performance. To mitigate this, practices such as random spot-checking or re-running a subset of inferences could be implemented to detect inconsistencies and ensure fairness.

$P_{\useworkflowIndex}$ \textbf{Direct Model Evaluation} \\ 
\emph{Rationale}: Leaderboard operators (\raisebox{-0.5mm}{\adjustbox{height=0.8\baselineskip,valign=m}{\includegraphics{figures/rq1/LO.png}}}) and/or external contributors (\raisebox{-0.5mm}{\adjustbox{height=0.8\baselineskip,valign=m}{\includegraphics{figures/rq1/EC.png}}}) submit their models (\raisebox{-0.5mm}{\adjustbox{height=0.8\baselineskip,valign=m}{\includegraphics{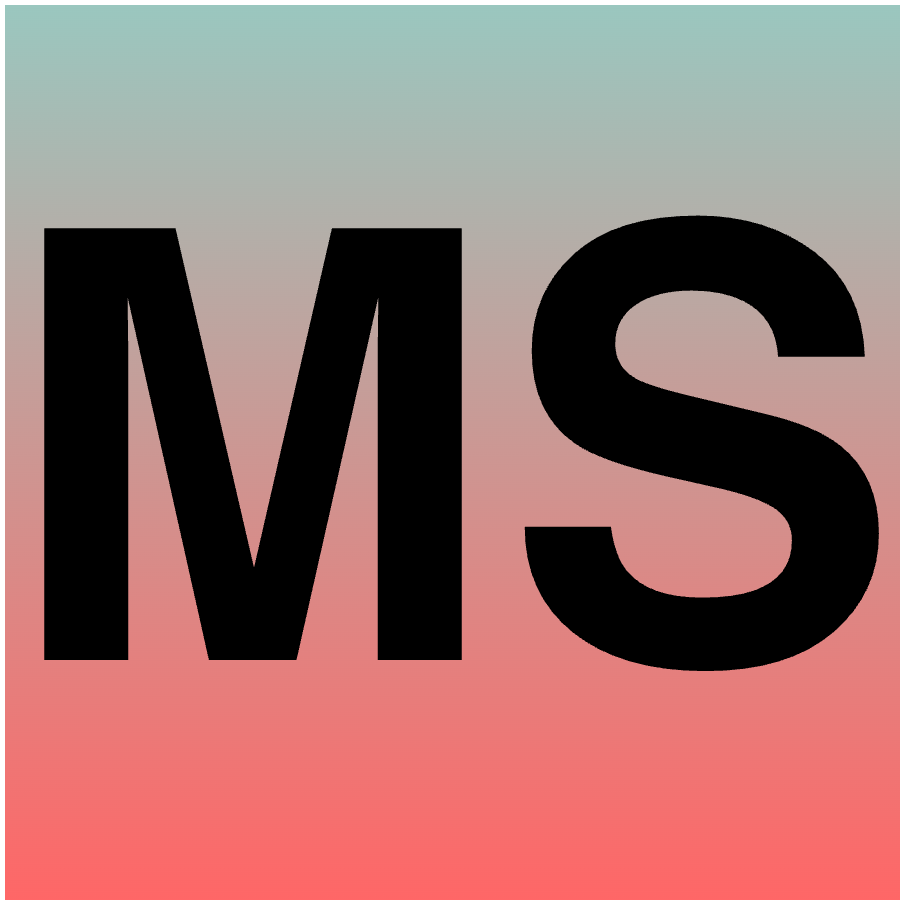}}}) through designated channels to the leaderboard. The models (\raisebox{-0.5mm}{\adjustbox{height=0.8\baselineskip,valign=m}{\includegraphics{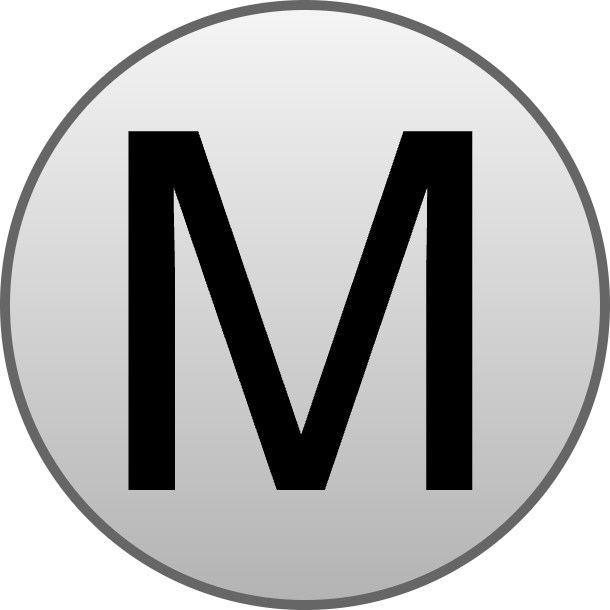}}}) usually include artifacts, such as repository URLs, APIs, binaries, and their configuration settings. Then these models are evaluated (\raisebox{-0.5mm}{\adjustbox{height=0.8\baselineskip,valign=m}{\includegraphics{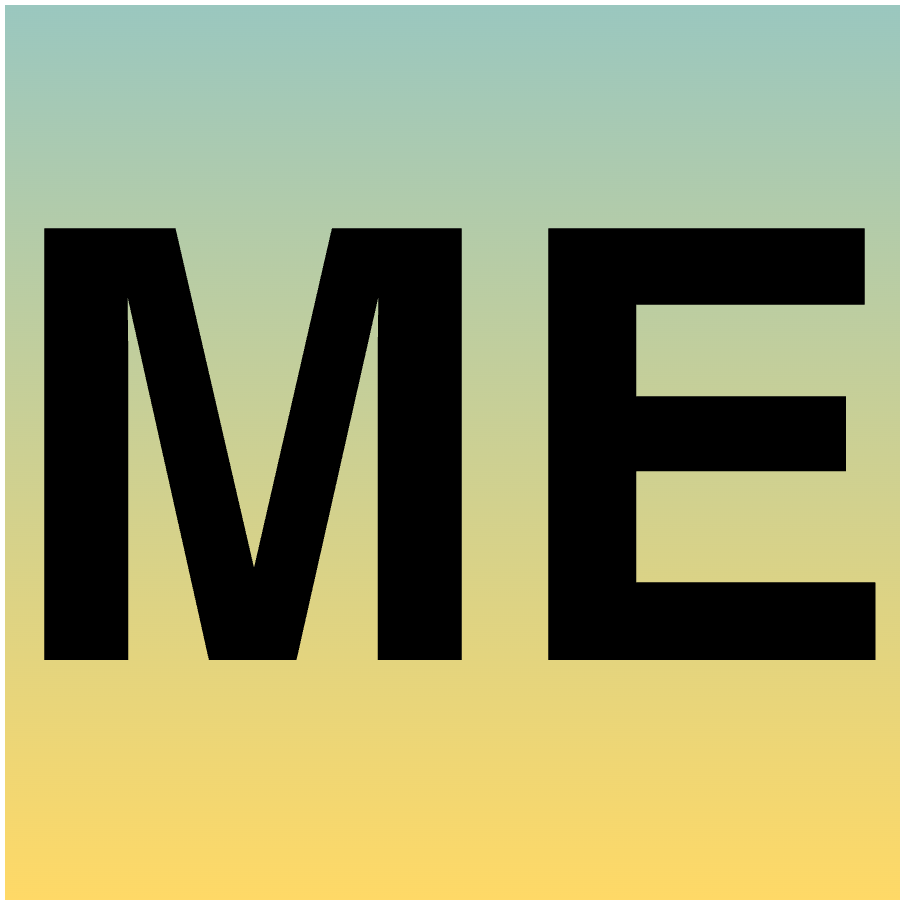}}}) by leaderboard operators or independent judges (\raisebox{-0.5mm}{\adjustbox{height=0.8\baselineskip,valign=m}{\includegraphics{figures/rq1/IJ.png}}}) directly based on either personal preferences or predefined benchmark protocol. Finally, the evaluation scores as records (\raisebox{-0.5mm}{\adjustbox{height=0.8\baselineskip,valign=m}{\includegraphics{figures/rq1/E.png}}}) are integrated (\raisebox{-0.5mm}{\adjustbox{height=0.8\baselineskip,valign=m}{\includegraphics{figures/rq1/RI.png}}}) into new or existing ranking dataframes (\raisebox{-0.5mm}{\adjustbox{height=0.8\baselineskip,valign=m}{\includegraphics{figures/rq1/R.png}}}) for future releases. The continuous nature of the record integration process (\raisebox{-0.5mm}{\adjustbox{height=0.8\baselineskip,valign=m}{\includegraphics{figures/rq1/CI.png}}}) ensures that the leaderboards remain current and relevant.  \\
\emph{Example}: \href{https://huggingface.co/spaces/NexaAIDev/domain_llm_leaderboard}{Domain LLM Leaderboard}, \href{https://llmleaderboard.goml.io}{LLM Use Case Leaderboard}, \href{https://opening-up-chatgpt.github.io}{Openness Leaderboard}. \\
\emph{Discussion}: Direct model evaluation is evaluator-driven and does not involve automated model inference. This approach manifests itself in two primary forms: community-driven evaluation and operator-driven evaluation. Community-driven methods, such as those used in the \href{https://llmleaderboard.goml.io}{LLM Use Case Leaderboard}, adopt a ``wisdom of the crowd'' approach where independent judges rank models based on personal preferences, similar to GitHub ``stars'' or Hugging Face ``likes''. While scalable and inclusive, this method introduces bias and challenges in ranking interpretability due to subjective opinions influenced by various expertise and criteria~\cite{borges2018s}. On the other hand, operator-driven evaluation, as seen in the \href{https://opening-up-chatgpt.github.io}{Openness Leaderboard}, ensures structured assessments of specific quality attributes, such as model openness~\cite{liesenfeld2023opening}, reducing subjectivity but limiting scalability due to the labor-intensive nature of manual evaluation. 

$P_{\useworkflowIndex}$ \textbf{Pointwise Model Evaluation}~\cite{xiong2024llava} \\ 
\emph{Rationale}: Leaderboard operators (\raisebox{-0.5mm}{\adjustbox{height=0.8\baselineskip,valign=m}{\includegraphics{figures/rq1/LO.png}}}) and/or external contributors (\raisebox{-0.5mm}{\adjustbox{height=0.8\baselineskip,valign=m}{\includegraphics{figures/rq1/EC.png}}}) submit their models (\raisebox{-0.5mm}{\adjustbox{height=0.8\baselineskip,valign=m}{\includegraphics{figures/rq1/MS.png}}}) through designated channels to the leaderboard. Then, leaderboard operators or independent judges (\raisebox{-0.5mm}{\adjustbox{height=0.8\baselineskip,valign=m}{\includegraphics{figures/rq1/IJ.png}}})—either AIs (\eg, MLLM-as-a-Judge~\cite{chen2024mllm}) or humans (\eg, \href{https://github.com/jeinlee1991/chinese-llm-benchmark}{Chinese Large Model Leaderboard})—perform inference (\raisebox{-0.5mm}{\adjustbox{height=0.8\baselineskip,valign=m}{\includegraphics{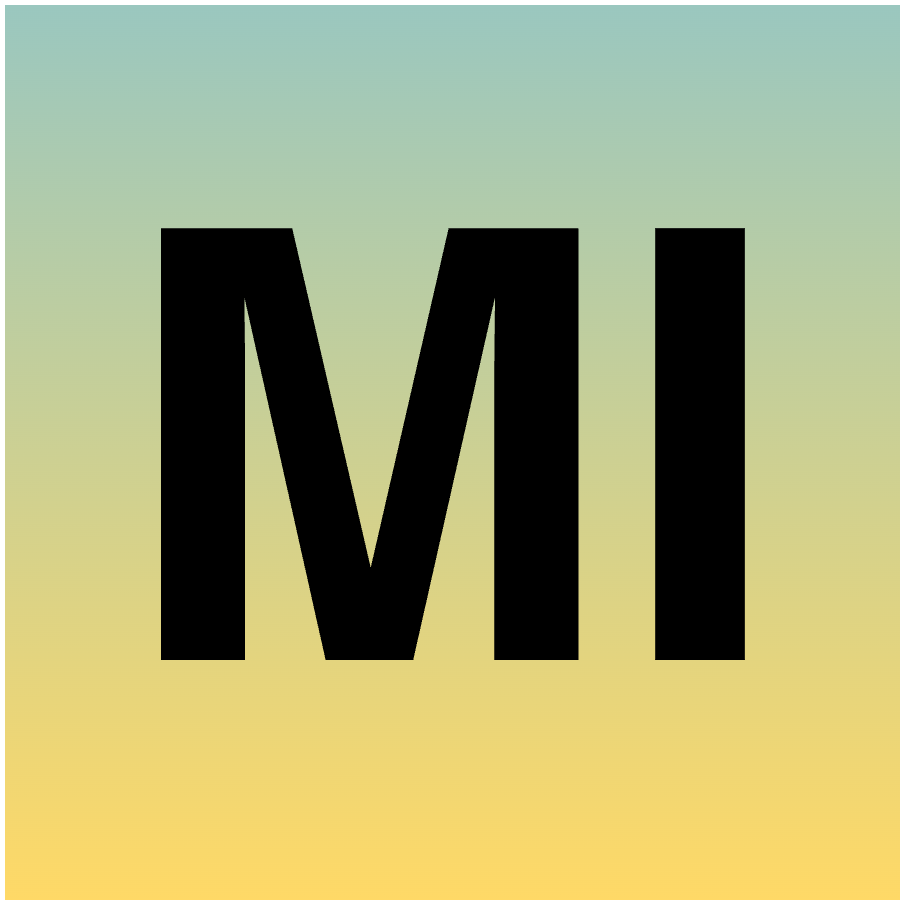}}}) on the benchmark test set using candidate models (\raisebox{-0.5mm}{\adjustbox{height=0.8\baselineskip,valign=m}{\includegraphics{figures/rq1/M.png}}}) to generate outputs (\raisebox{-0.5mm}{\adjustbox{height=0.8\baselineskip,valign=m}{\includegraphics{figures/rq1/O.png}}}). Subsequently, these outputs are evaluated (\raisebox{-0.5mm}{\adjustbox{height=0.8\baselineskip,valign=m}{\includegraphics{figures/rq1/OE.png}}}) against the benchmark ground truths. If ground truths are not available, independent judges act as reviewers to directly assign scores following predefined benchmark protocols. Finally, the evaluation scores as records (\raisebox{-0.5mm}{\adjustbox{height=0.8\baselineskip,valign=m}{\includegraphics{figures/rq1/E.png}}}) are integrated (\raisebox{-0.5mm}{\adjustbox{height=0.8\baselineskip,valign=m}{\includegraphics{figures/rq1/RI.png}}}) into new or existing ranking dataframes (\raisebox{-0.5mm}{\adjustbox{height=0.8\baselineskip,valign=m}{\includegraphics{figures/rq1/R.png}}}) for future releases. The continuous nature of record integration process (\raisebox{-0.5mm}{\adjustbox{height=0.8\baselineskip,valign=m}{\includegraphics{figures/rq1/CI.png}}}) ensures leaderboards to remain current and relevant. \\
\emph{Example}: \href{https://evalplus.github.io/leaderboard.html}{EvalPlus}, \href{https://flageval.baai.ac.cn/\#/leaderboard}{FlagEval}, \href{https://crfm.stanford.edu/helm}{HELM leaderboards}. \\
\emph{Discussion}: Pointwise evaluation follows a structured approach where models are centrally evaluated against predefined benchmarks on the leaderboard, ensuring consistent and repeatable results if best practices are followed\footnote{\url{https://www.latent.space/p/benchmarks-201}}. This method is similar to a regression task~\cite{wang2024helpsteer2}, where the goal is to assign scores to model responses based on their alignment with ground truths, or predefined evaluation criteria in the absence of ground truths (\eg, evaluations by independent judges). This setup provides detailed, pointwise insights into model performance, making it reliable for assessing specific capabilities. However, limiting evaluations to a set of benchmarks can narrow the scope, potentially overlooking model performance in real-world scenarios where data diverge from the test set~\cite{zheng2024judging,lu2018learning,maslej2024artificialintelligenceindexreport}. Risks such as benchmark leakage can further compromise evaluations if developers overfit models to public benchmarks\footnote{\url{https://www.aisnakeoil.com/p/ai-leaderboards-are-no-longer-useful}}. From a scalability standpoint, pointwise evaluations can be resource-intensive, particularly when handling large datasets or computationally expensive benchmarks, such as HELM Classics~\cite{polo2024tinybenchmarks}. For smaller leaderboards, the costs of running extensive test suites and maintaining computational infrastructure to accommodate diverse model environments can pose significant challenges\footnote{\url{https://huggingface.co/spaces/open-llm-leaderboard/open_llm_leaderboard/discussions/801}}.

$P_{\useworkflowIndex}$ \textbf{Pairwise Model Evaluation} \\
\emph{Rationale}: Leaderboard operators (\raisebox{-0.5mm}{\adjustbox{height=0.8\baselineskip,valign=m}{\includegraphics{figures/rq1/LO.png}}}) and/or external contributors (\raisebox{-0.5mm}{\adjustbox{height=0.8\baselineskip,valign=m}{\includegraphics{figures/rq1/EC.png}}}) submit their models (\raisebox{-0.5mm}{\adjustbox{height=0.8\baselineskip,valign=m}{\includegraphics{figures/rq1/MS.png}}}) through designated channels to a leaderboard. Once submitted, the models (\raisebox{-0.5mm}{\adjustbox{height=0.8\baselineskip,valign=m}{\includegraphics{figures/rq1/M.png}}}) undergo a ``pairwise comparison'' process, where either independent judges (\raisebox{-0.5mm}{\adjustbox{height=0.8\baselineskip,valign=m}{\includegraphics{figures/rq1/IJ.png}}})—AIs (\eg, Auto-Arena) or humans (\eg, Chatbot Arena~\cite{chiang2024chatbot})—or leaderboard operators act as reviewers. These judges perform inferences (\raisebox{-0.5mm}{\adjustbox{height=0.8\baselineskip,valign=m}{\includegraphics{figures/rq1/MI.png}}}) on two models using predefined or arbitrary inquiries, then conduct blind tests to compare the outputs (\raisebox{-0.5mm}{\adjustbox{height=0.8\baselineskip,valign=m}{\includegraphics{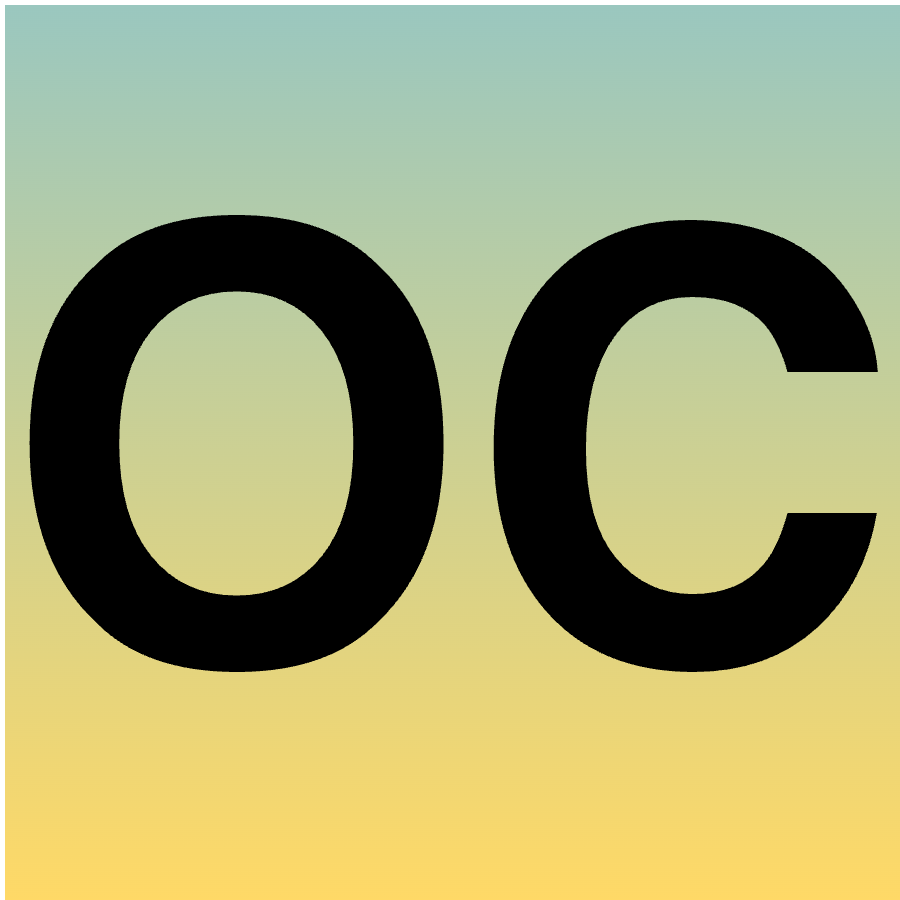}}})—either between the candidate models (\eg, Chatbot Arena~\cite{chiang2024chatbot}) or against a baseline model (\eg, Auto-J~\cite{li2023generative}). Usually, judges vote for the preferred model or declare a tie if neither model stands out. Votes are used to generate relative metric scores, such as Elo ratings (\eg, Chatbot Arena~\cite{chiang2024chatbot}) or human alignment rates (\eg, Auto-J). Finally, the evaluation scores as records (\raisebox{-0.5mm}{\adjustbox{height=0.8\baselineskip,valign=m}{\includegraphics{figures/rq1/E.png}}}) are integrated (\raisebox{-0.5mm}{\adjustbox{height=0.8\baselineskip,valign=m}{\includegraphics{figures/rq1/RI.png}}}) into ranking dataframes (\raisebox{-0.5mm}{\adjustbox{height=0.8\baselineskip,valign=m}{\includegraphics{figures/rq1/R.png}}}) for future releases. The continuous nature of record integration process (\raisebox{-0.5mm}{\adjustbox{height=0.8\baselineskip,valign=m}{\includegraphics{figures/rq1/CI.png}}}) ensures leaderboards to remain current and relevant. \\
\emph{Example}: \href{https://lmarena.ai/?leaderboard}{Chatbot Arena}, \href{https://llm-council.com}{Language Model Council}, \href{https://huggingface.co/spaces/HishamYahya/ZeroSumEval_Leaderboard}{ZeroSumEval}. \\
\emph{Discussion}: Pairwise evaluation employs a comparison-based approach, assessing models against each other through blind tests. This method, grounded in the Bradley-Terry model~\cite{bradley1952rank}, leverages preference-based feedback to highlight contrasts between selected and rejected responses. It effectively models real-world user preferences by generating relative scores, which streamline the evaluation process and reduce the cognitive load of human judges, thus improving engagement. The blind evaluation process also helps mitigate bias, leading to more balanced assessments\footnote{\url{https://x.com/DrJimFan/status/1833160432833716715}}. However, some evaluation metrics, such as Elo ratings~\cite{hvattum2010using}—originally designed to estimate player skill based on pairwise comparisons—assume that the skill of the model remains constant over time, which may fail to reflect continuous advancements in FMs~\cite{volkovs2012flexible}. From a scalability standpoint, pairwise evaluation encounters challenges as the number of models grows, with each new model submitted exponentially increasing the required comparisons to sustain robust rankings\footnote{\url{https://bryanyzhu.github.io/posts/2024-06-20-elo-part1}}. A notable solution to these scalability issues is the Decentralized Arena\footnote{\url{https://de-arena.maitrix.org}} where all models participating in the evaluation process also serve as judges for other models within a sliding window. Furthermore, pairwise comparisons often face reproducibility challenges, as rankings can fluctuate due to subjective factors, including the timing of evaluations, the varying difficulty of queries presented to models, and biases influenced by the length or formatting of model outputs\footnote{\url{https://www.latent.space/p/lmarena}}.

Table~\ref{tab:platform workflow} provides a breakdown of the distribution of the five workflow patterns across different sources. Overall, $P_1$ (External Evaluation Integration) is the most prevalent workflow pattern, accounting for $61.63\%$ ($644/1,045$) of the collected leaderboards. This is followed by $P_4$ (Pointwise Model Evaluation) at $41.15\%$, $P_2$ (Model Output Evaluation) at $9.09\%$, $P_5$ (Pairwise Model Evaluation) at $2.58\%$, and $P_3$ (Direct Model Evaluation) at $0.38\%$. For $P_1$, PWC leaderboards predominantly follow this pattern ($99.65\%$), making it the most common workflow on specific platforms ($85.45\%$). $P_2$ is most frequently used on GitHub ($37.61\%$), followed closely by independent platforms ($36.75\%$). $P_3$, while rare, is most prevalent in HF Spaces ($50\%$), followed equally by GitHub and independent platforms (both $25\%$). $P_4$ is most common on GitHub ($43.65\%$), followed by HF Spaces ($31.35\%$) and independent platforms ($23.81\%$). $P_5$ is most often used on HF Spaces, where it accounts for $54.55\%$ leaderboards, followed by independent platforms ($24.24\%$) and GitHub ($21.21\%$). Furthermore, we observe that $14.35\%$ ($150/1,045$) leaderboards adopt multiple workflow patterns to accommodate different evaluation objectives. Among these, $P_2$ (Model Output Evaluation) and $P_4$ (Pointwise Model Evaluation) stand out as the most common pairing, accounting for $44.67\%$ ($67/150$). An example of this pairing is the \href{https://huggingface.co/spaces/wzxii/Memorization-or-Generation-of-Big-Code-Models-Leaderboard}{Memorization or Generation of Big Code Models Leaderboard}.

\begin{table}[t]
\centering
\caption{The distribution of FM leaderboards with specific workflow patterns across different sources.}
\label{tab:platform workflow}
\begin{NiceTabular}{l|C{1cm}C{1cm}C{1cm}C{1cm}C{1cm}}
& \multicolumn{5}{c}{\textbf{Workflow Pattern}} \\ 
\cline{2-6}
\textbf{Source} & $P_1$ & $P_2$ & $P_3$ & $P_4$ & $P_5$ \\ \midrule
GitHub & 6.52\% (43/660) & 37.61\% (44/117) & 25.00\% (1/4) & 43.65\% (220/504) & 21.21\% (7/33) \\ \midrule
HF Spaces & 5.00\% (33/660) & 23.08\% (27/117) & 50.00\% (2/4) & 31.35\% (158/504) & 54.55\% (18/33) \\ \midrule
\makecell{independent \\ platform} & 2.73\% (18/660) & 36.75\% (43/117) & 25.00\% (1/4) & 23.81\% (120/504) & 24.24\% (8/33) \\ \midrule
PWC & 85.45\% (564/660) & 1.71\% (2/117) & 0.00\% (0/4) & 0.40\% (2/504) & 0.00\% (0/33) \\ \midrule
\makecell{spreadsheet \\ platform} & 0.30\% (2/660) & 0.85\% (1/117) & 0.00\% (0/4) & 0.79\% (4/504) & 0.00\% (0/33) \\ \midrule
Overall & 61.63\% (644/1045) & 9.09\% (95/1045) & 0.38\% (4/1045) & 41.15\% (430/1045) & 2.58\% (27/1045) \\
\end{NiceTabular}
\end{table}

\subsection{Domain Model}
\label{sec:rq1-results:domain-model}

\begin{figure}
\centering
\includegraphics[width=\linewidth]{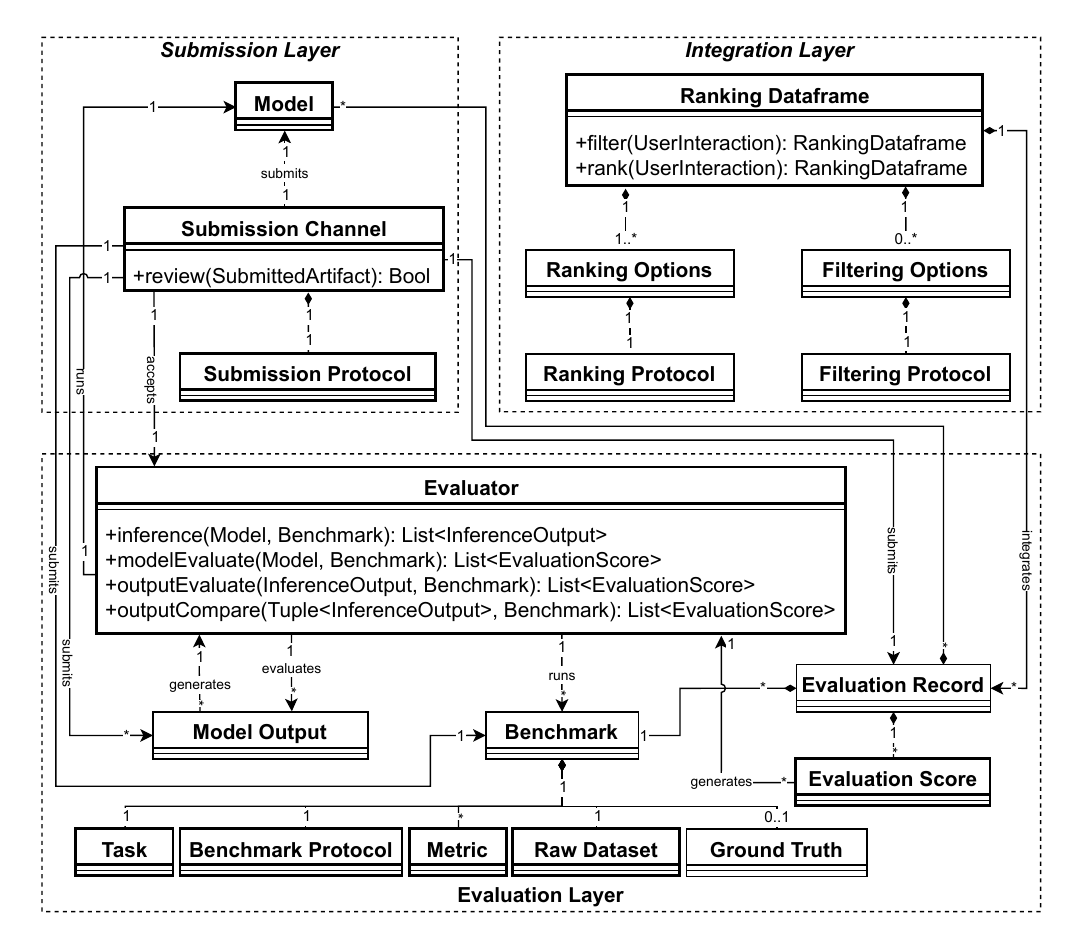}
\caption{The domain model of LBOps leveraged by the five identified workflow patterns. The level of adherence depends on the specific pattern and leaderboard.}
\label{fig:domain model}
\centering
\end{figure}

\begin{figure*}
\centering
\includegraphics[width=0.75\linewidth]{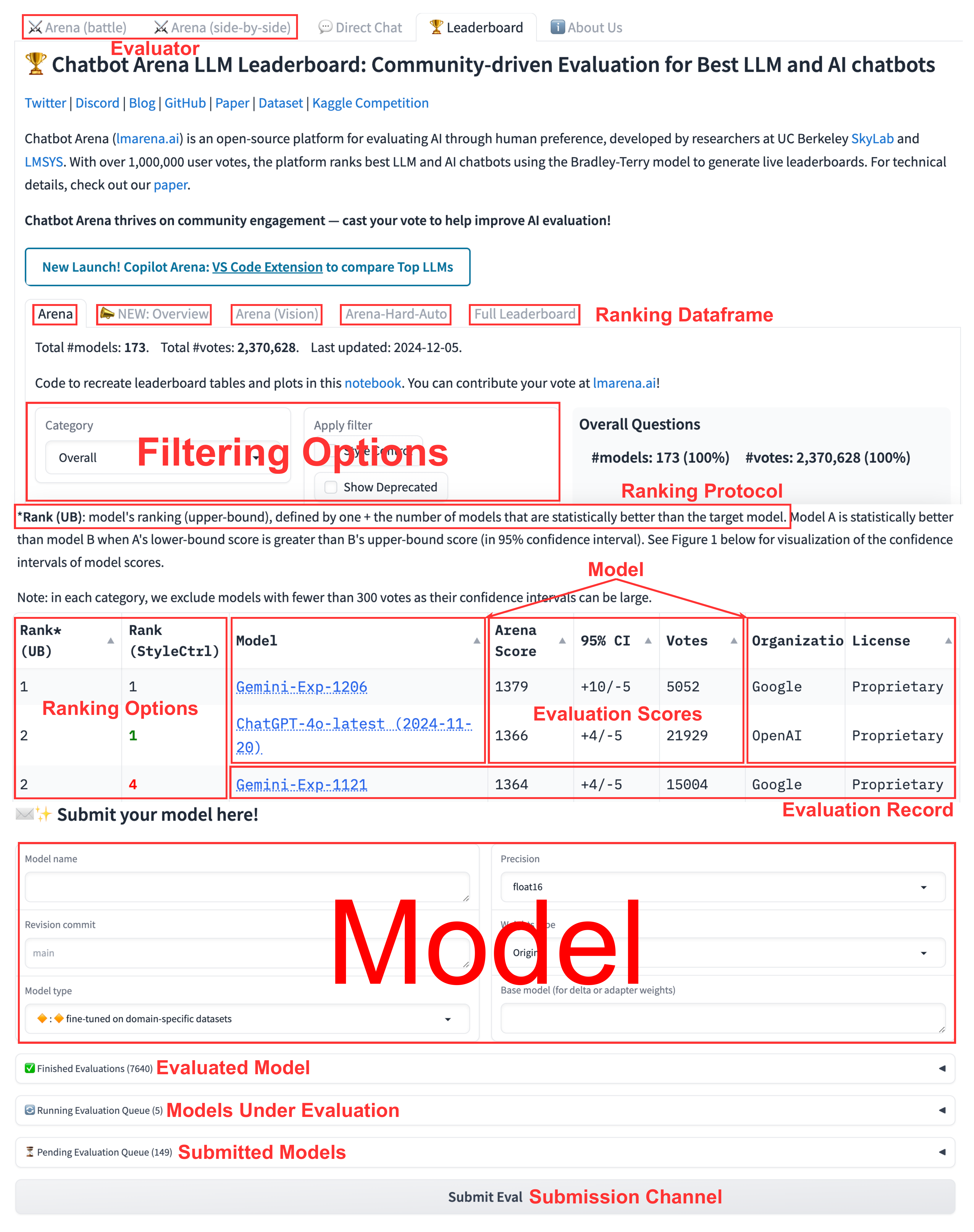}
\caption{Collage screenshot of leaderboard components using elements from the \href{https://lmarena.ai/?leaderboard}{Chatbot Arena} and \href{https://huggingface.co/spaces/open-llm-leaderboard/open_llm_leaderboard}{Open LLM Leaderboard}.}
\label{fig:leaderboard structure}
\centering
\end{figure*}

Figure~\ref{fig:domain model} presents the domain model of LBOps workflows, offering a comprehensive structured view of its components and their interrelationships. This model is structured into three layers (submission, evaluation, and integration), aligning with the corresponding phases of the workflow patterns shown in Figure~\ref{fig:workflow pattern}. Complementing this, Figure~\ref{fig:leaderboard structure} provides a visual representation of the common components of the domain model, highlighting elements that are explicitly visible in typical leaderboard GUIs. To aid in understanding, the following section provides a detailed explanation. For clarity, we omit common metadata, including the release date, version number, and stakeholder name, for each component in the domain model.

\emph{Submission channel} serves as the dedicated avenue for users to submit \textit{model}, \textit{model output}, \textit{evaluation records} or even \textit{benchmarks}, to leaderboards. The \textit{submission channels} involve \href{https://github.com/rowanz/hellaswag/tree/master/hellaswag_models\#submitting-to-the-leaderboard}{emails}, \href{https://github.com/tatsu-lab/alpaca_eval?tab=readme-ov-file\#contributing-a-model}{pull requests}, \href{https://github.com/ray-project/llmperf-leaderboard?tab=readme-ov-file\#feedback}{issue trackers}, \href{https://github.com/paperswithcode/model-index}{model cards}, \href{https://huggingface.co/spaces/open-llm-leaderboard/open_llm_leaderboard}{upload portal}, and/or \href{https://huggingface.co/spaces/Cognitive-Lab/indic_llm_leaderboard}{API calling}. This channel may be private to only leaderboard operators, as seen in the case of \href{https://crfm.stanford.edu/helm}{HELM leaderboards}. The model submitted via such channels typically includes vital details about the model, such as its name, publisher, release date, type (\eg, fine-tuned or base model), parameter count, publication name, repository linkage, and API token, which provides crucial context for understanding the model's characteristics and provenance.

Upon submission, the leaderboard operators might review the provided artifacts, \eg, \textit{model}, and verify its adherence to the submission protocol. For example, HF users must adhere to the metadata format specified in their model cards when submitting a model to the PWC leaderboards. Once a \textit{model} passes the review stage, it is then evaluated against predefined \textit{benchmarks} using the \textit{evaluator} (detailed below). If the \textit{model} does not meet the specified requirements during the review process, it can be rejected or excluded from the pending evaluation queue. This review protocol ensures that only models adhering to the submission protocol advance to the subsequent stages of evaluation and comparison.

A \emph{benchmark} is an evaluation framework to assess the performance of ML models~\cite{denton2020bringing,dueben2022challenges,thiyagalingam2022scientific}. A comprehensive ML benchmark typically comprises five key components: the \textit{task}, \textit{protocol}, \textit{raw dataset}, \textit{ground truth}, and \textit{metrics}. The \textit{task} defines the specific goals or challenges that ML models aim to achieve or address, while the \textit{protocol} establishes a set of rules and guidelines for the evaluation process. The \textit{raw dataset} comprises structured or unstructured data samples that serve as the basis for model evaluation. Complementing the raw dataset, the \textit{ground truth} is a collection of descriptive labels, often referred to as \href{https://stats.stackexchange.com/questions/333446/what-does-the-term-gold-label-refer-to-in-the-context-of-semi-supervised-class}{``gold labels''}. These labels represent reliable and referential outcomes for given inputs and can be acquired either through the work of human annotators or via FM output. Finally, \textit{metrics} offer quantitative measures to evaluate model performance, enabling researchers to identify areas for improvement and track progress within the field.

\emph{Evaluator} refers to a suite of software tools and frameworks designed to execute FM evaluations against predefined benchmarks~\cite{bommasani2023holistic}. However, in specific contexts, it may also encompass human or AI judges~\cite{zheng2024judging}. A notable example is \href{https://github.com/lm-sys/FastChat}{FastChat}, which serves as an evaluator for \href{https://lmarena.ai/?leaderboard}{Chatbot Arena}. The evaluator follows a structured process to assess submitted models, consisting of two main approaches: a sequential process involving model inference and output evaluation, or a unified process of direct model evaluation. Typically, drawing from the evaluation queue of pending models, the evaluator retrieves the top-most \textit{model} and commences the evaluation pipeline. Initially, the evaluator performs inferences on the test \textit{raw dataset} while adhering to the predefined benchmark \textit{protocol}. For direct evaluation, leaderboard operators or independent judges assign scores or cast votes for specific models based on personal preferences or predefined protocols. Subsequently, the evaluator evaluates the \textit{model output} against the \textit{ground truth} to obtain the evaluation \textit{metric} scores. For output comparison, models are evaluated against each other, incorporating judge feedback and generating Elo-like scores. Upon completion of these steps, the evaluator combines the model with \textit{evaluation scores}, generating an \textit{evaluation record}. These records are then ready to be integrated into new or existing \textit{ranking dataframes}, facilitating further analysis and comparison of different models.

\emph{Ranking dataframe} is a specialized dataframe that contains \textit{evaluation records}, along with \textit{ranking and filtering options} tailored to specific protocols (detailed below). A single leaderboard may contain multiple \textit{ranking dataframes}, each serving a unique purpose in analyzing and comparing \textit{model} performance. For example, as of \collectdate, there are $5$ ranking dataframes on the \href{https://lmarena.ai/?leaderboard}{Chatbot Arena}: ``Arena'', ``Overview'', ``Arena (Vision)'', ``Arena-Hard-Auto'', ``Full Leaderboard''. \textit{Ranking dataframes} can be presented in various formats, including tables (\eg, \href{https://tatsu-lab.github.io/alpaca_eval}{regular table}, \href{https://huggingface.co/spaces/gsaivinay/open_llm_leaderboard}{rankable table}, \href{https://github.com/MikeGu721/XiezhiBenchmark}{table screenshot}), figures (\eg, \href{https://leaderboard.tabbyml.com}{bar chart}, \href{https://artificialanalysis.ai/leaderboards/models}{box plot}, \href{https://videoniah.github.io}{heat map}, \href{https://xwang.dev/mint-bench}{line chart}, \href{https://osu-nlp-group.github.io/TravelPlanner}{pie chart}, \href{https://huggingface.co/spaces/BramVanroy/open_dutch_llm_leaderboard}{radar chart}, \href{https://huggingface.co/spaces/ml-energy/leaderboard}{scatter plot}, \href{https://huggingface.co/spaces/ramiroluo/LLMHallucination_Leaderboard}{sortable bar chart}, and even \href{https://github.com/AINativeLab/gptstore-data-backup}{sequential text} enabling a customized user experience for performance comparison and highlighting crucial insights. Specialized toolkits, such as \href{https://huggingface.co/spaces/dimbyTa/open-llm-leaderboard-viz}{Open LLM Leaderboard Viz}, can visualize raw leaderboards in different formats.

\textit{Ranking options} determine how models are ranked based on the \textit{ranking protocol}, which establishes rules for preferring one evaluated model over another. These protocols ensure that when a model is preferred to (or beats) its counterparts, it holds a higher position within the leaderboard rankings. In the Chatbot Arena example, the default ranking dataframe employs the metric ``bound score'' to rank the evaluated models, defined as ``one + the number of models that are statistically better than the target model''. Some ranking dataframes allow users to select their preferred metrics for ranking by interacting with the metric names, providing a dynamic and customizable comparison experience in a rankable display format.

\textit{Filtering options} determine which evaluation records are included or excluded from specific ranking dataframes, guided by a \textit{filtering protocol}. These protocols account for factors such as model type, tasks, languages, and other relevant characteristics to ensure the displayed records meet the user's needs or preferences. In the Chatbot Arena example, users can switch between more than $20$ ranking frames using predefined filtering options, such as ``Overall'', ``Coding'', ``French'', \etc, by interacting with the ``Category'' tab. In some leaderboards, such as \href{https://huggingface.co/spaces/open-llm-leaderboard/open_llm_leaderboard}{Open LLM Leaderboard}, users can customize ad hoc ranking dataframes by entering model keywords into the search box. 

\begin{mybox}{$RQ_1$ Summary}
\begin{itemize}[leftmargin=*]
    \item We identify five different workflow patterns that represent the state-of-the-art of LBOps. We also derive the domain model that encapsulates the essential components, attributes, and interactions inherent to LBOps.
    \item We find that $P_1$ (External Evaluation Integration) is the most prevalent workflow pattern in LBOps, followed by $P_4$ (Pointwise Model Evaluation), $P_2$ (Model Output Evaluation), $P_5$ (Pairwise Model Evaluation), and $P_3$ (Direct Model Evaluation).
\end{itemize}
\end{mybox}

\section{$RQ_2$ Results: Leaderboard Smells}
\label{sec:rq2-results}

This section presents our identified leaderboard smell and their associated components in our domain model, as listed in Table~\ref{tab:leaderboad smell} and highlighted in Figure~\ref{fig:domain model}. The unchecked cells in the table indicate that no evidence of a specific smell was found within the scope of this study for the corresponding component. However, this absence of evidence does not rule out the possibility that such smells may manifest in other components under different conditions.

\begin{table}[t]
\centering
\caption{The distribution of smells occurring within different leaderboard components.}
\label{tab:leaderboad smell}
    \begin{NiceTabular}{l|lllllllll}
         & \multicolumn{9}{c}{\textbf{Leaderboard Component}} \\ 
         \cline{2-10}
         \textbf{Smell Type} & \rotatebox{90}{Benchmark Metric} & \rotatebox{90}{Benchmark Protocol} & \rotatebox{90}{Benchmark Raw Dataset} & \rotatebox{90}{Benchmark Task} & \rotatebox{90}{Evaluator} & \rotatebox{90}{Evaluation Score} & \rotatebox{90}{Model} & \rotatebox{90}{Ranking Dataframe} & \rotatebox{90}{Submission Channel/Protocol} \\ \midrule
        Confusing Entity & \checkmark & \checkmark & \checkmark & \checkmark &  & \checkmark & \checkmark & \checkmark & \checkmark \\ \midrule
        Deprecated Entity &  &  &  & \checkmark &  &  &  & \checkmark &  \\ \midrule
        Inaccessible Entity &  &  & \checkmark &  &  & \checkmark & \checkmark & \checkmark & \checkmark \\ \midrule
        Misdisplayed Entity & \checkmark & \checkmark &  &  &  & \checkmark &  & \checkmark &  \\ \midrule
        Mismatched Entity & \checkmark & \checkmark & \checkmark & \checkmark &  & \checkmark & \checkmark & \checkmark &  \\ \midrule
        Missing Entity & \checkmark & \checkmark & \checkmark &  &  & \checkmark & \checkmark & \checkmark & \checkmark \\ \midrule
        Redundant Entity &  &  &  & \checkmark &  & \checkmark & \checkmark & \checkmark &  \\ \midrule
        Unresponsive Entity &  &  &  &  & \checkmark &  &  & \checkmark & \checkmark \\
    \end{NiceTabular}
\end{table}

In the next sections, we explain each type of leaderboard smell, providing a relevant example to illustrate its occurrence within the context of various leaderboard components. 

\subsection{Confusing Entity Smell}
This smell refers to vaguely defined or specified entities on the leaderboard. Confusing entities can cause misinterpretation, misinformed decisions, and decreased trust. \\
\hrule \noindent \\
\emph{Component}: Benchmark Metric \\
\emph{Example}: The metrics used for the BiPaR~\cite{jing2019bipar} and \href{https://github.com/junzeng-pluto/ChineseSquad}{ChineseSquad} benchmarks are not clearly defined on the OpenEval leaderboard\footnote{\url{https://github.com/tjunlp-lab/Awesome-LLMs-Evaluation-Papers/issues/21}}. \\
\hrule \noindent \\
\emph{Component}: Benchmark Protocol \\
\emph{Example}: The Coding LLMs Leaderboard lacks clarity on key aspects of the evaluation protocol, such as the methodology used to derive evaluation scores\footnote{\url{https://github.com/TabbyML/tabby/issues/1369}}. \\
\hrule \noindent \\
\emph{Component}: Benchmark Raw Dataset \\
\emph{Example}: The Nexus Function Calling Leaderboard failed to specify the benchmark datasets\footnote{\url{https://huggingface.co/spaces/Nexusflow/Nexus_Function_Calling_Leaderboard/discussions/3}}. \\
\hrule \noindent \\
\emph{Component}: Benchmark Task \\
\emph{Example}: The number of scenarios listed on the ``Scenario'' page does not match the count displayed on the main page of the HELM Classic leaderboard\footnote{\url{https://github.com/stanford-crfm/helm/issues/2028}}. \\
\hrule \noindent \\
\emph{Component}: Evaluation Score \\
\emph{Example}: The SeaEval leaderboard listed hundreds of performance scores with the impossible value of ``$-1$''\footnote{\url{https://huggingface.co/spaces/SeaEval/SeaEval_Leaderboard/discussions/2}}. \\
\hrule \noindent \\
\emph{Component}: Model \\
\emph{Example}: The meaning of greyed-out models is not defined in the HELM Classic leaderboard\footnote{\url{https://github.com/stanford-crfm/helm/issues/2386}}. \\
\hrule \noindent \\
\emph{Component}: Ranking Dataframe \\
\emph{Example}: The CanAiCode Leaderboard did not explain the ``best result only'' filtering option\footnote{\url{https://huggingface.co/spaces/mike-ravkine/can-ai-code-results/discussions/6}}. \\
\hrule \noindent \\
\emph{Component}: Submission Channel/Protocol \\
\emph{Example}: The SWE-bench leaderboard did not provide specific details on how to submit new evaluations\footnote{\url{https://github.com/princeton-nlp/SWE-bench/issues/46}}. \\

\subsection{Deprecated Entity Smell}
This smell refers to obsolete entities on the leaderboard, which still serve as archives but no longer contribute meaningfully to SOTA model comparisons. Deprecated entities can cause inaccurate rankings, misinformed decisions, and extra maintenance overhead. \\
\hrule \noindent \\
\emph{Component}: Benchmark Task \\
\emph{Example}: The ``junior-dev'' benchmark was deprecated in the CanAiCode Leaderboard\footnote{\url{https://huggingface.co/spaces/mike-ravkine/can-ai-code-results/discussions/3}}. \\
\hrule \noindent \\
\emph{Component}: Ranking Dataframe \\
\emph{Example}: The C-Eval leaderboard's GitHub evaluations are currently out of sync with its independent platform's ones\footnote{\url{https://github.com/hkust-nlp/ceval/issues/76}}. \\

\subsection{Inaccessible Entity Smell} 
This smell refers to entities on the leaderboard that are expected to be accessible online but remain unavailable for an extended period due to networking issues. Inaccessible entities can cause a lack of transparency, concerns about data integrity, imbalanced competition, and reduced user engagement. \\
\hrule \noindent \\
\emph{Component}: Benchmark Raw Dataset \\
\emph{Example}: The MS MARCO dataset~\cite{bajaj2016ms} was inaccessible from the MTEB leaderboard download portal for about three months\footnote{\url{https://github.com/embeddings-benchmark/mteb/issues/186}}. \\
\hrule \noindent \\
\emph{Component}: Evaluation Score \\
\emph{Example}: Most evaluations are inaccessible on the OPUS-MT Dashboard, as it can only display a maximum of 50 records in its ranking dataframes for any specified benchmark\footnote{\url{https://github.com/Helsinki-NLP/OPUS-MT-dashboard/issues/2}}. \\
\hrule \noindent \\
\emph{Component}: Model \\
\emph{Example}: Users are forbidden to access the models listed on the OpenVLM Leaderboard for unknown reasons\footnote{\url{https://huggingface.co/spaces/opencompass/open_vlm_leaderboard/discussions/6}}. \\
\hrule \noindent \\
\emph{Component}: Ranking Dataframe \\
\emph{Example}: The Safety Prompts leaderboard was inaccessible to external users for more than a week due to restrictions limited to the campus intranet\footnote{\url{https://github.com/thu-coai/Safety-Prompts/issues/18}}. \\
\hrule \noindent \\
\emph{Component}: Submission Channel/Protocol \\
\emph{Example}: Submission channels were inaccessible on the BookSQL leaderboard due to an incorrect link\footnote{\url{https://github.com/Exploration-Lab/BookSQL/issues/5}}. \\

\subsection{Misdisplayed Entity Smell}
This smell refers to entities on the leaderboard that are either incorrectly displayed or formatted. Misdisplayed entities can cause user frustration, decreased trust, and misinformed decisions. \\
\hrule \noindent \\
\emph{Component}: Benchmark Metric \\
\emph{Example}: In the ``Image Classification on ImageNet'' leaderboard, specific metrics, such as ``Hardware Burden'' and ``Operations per network pass'', were present in the HTML webpage source but could not be found in the actual ranking dataframes\footnote{\url{https://github.com/paperswithcode/sota-extractor/issues/25}}. \\
\hrule \noindent \\
\emph{Component}: Benchmark Protocol \\
\emph{Example}: The repeated inclusion of ``num\_instances=10'' in all ``synthetic efficiency'' ranking dataframe names on the HELM Classic leaderboard adds unnecessary clutter and complicates navigation\footnote{\url{https://github.com/stanford-crfm/helm/issues/2205}}. \\
\hrule \noindent \\
\emph{Component}: Evaluation Score \\
\emph{Example}: When filtering options are applied on the LLM Safety Leaderboard, models including ``anthropic/claude-2.0'' and ``openai/gpt-3.5-turbo-0301'' disappear, and unselecting the filters does not restore them\footnote{\url{https://huggingface.co/spaces/AI-Secure/llm-trustworthy-leaderboard/discussions/5}}. \\
\hrule \noindent \\
\emph{Component}: Ranking Dataframe \\
\emph{Example}: The Q-Bench leaderboard had a misaligned layout in its ``Overall Leaderboards'' ranking dataframe for over four months\footnote{\url{https://github.com/Q-Future/Q-Bench/issues/11}}. \\

\subsection{Mismatched Entity Smell}
This smell refers to the discrepancies between the claims made by the developers and the actual performance or attributes of the entities displayed on the leaderboard. Mismatched entities can cause user confusion, reduced reliability, and misinformed decisions. \\
\hrule \noindent \\
\emph{Component}: Benchmark Metric \\
\emph{Example}: There was a discrepancy between the number of benchmark metrics stated on the website and the actual metrics used in the HELM Classic leaderboard\footnote{\url{https://github.com/stanford-crfm/helm/issues/2034}}. \\ 
\hrule \noindent \\
\emph{Component}: Benchmark Protocol \\
\emph{Example}: The LLM-Leaderboard evaluation scores are based on 10-shot settings, not 0-shot, as verified by the MosaicML LLM Foundry's evaluation scripts\footnote{\url{https://github.com/LudwigStumpp/llm-leaderboard/issues/9}}. \\ 
\hrule \noindent \\
\emph{Component}: Benchmark Raw Dataset \\
\emph{Example}: The dataset name on the MTEB leaderboard did not match the actual dataset\footnote{\url{https://github.com/embeddings-benchmark/mteb/issues/191}}. \\ 
\hrule \noindent \\
\emph{Component}: Benchmark Task \\
\emph{Example}: The ``HumanEval'' leaderboard was mislabeled as ``Image Classification'', as HumanEval~\cite{chen2021evaluating} is designed for code generation tasks\footnote{\url{https://web.archive.org/web/20221205184917/https://paperswithcode.com/sota/image-classification-on-humaneval}}. \\ 
\hrule \noindent \\
\emph{Component}: Evaluation Score \\
\emph{Example}: The YALL leaderboard incorrectly shows the ``google/gemma-2b-it'' model repository with an impossible value of ``$-1$'' likes, instead of the correct count of $478$\footnote{\url{https://huggingface.co/spaces/mlabonne/Yet_Another_LLM_Leaderboard/discussions/11}}. \\ 
\hrule \noindent \\
\emph{Component}: Model \\
\emph{Example}: Hundreds of models on the Open PL LLM Leaderboard are listed with an impossible parameter count of $0$\footnote{\url{https://huggingface.co/spaces/speakleash/open_pl_llm_leaderboard/discussions/1}}. \\ 
\hrule \noindent \\
\emph{Component}: Ranking Dataframe \\
\emph{Example}: The GPTEval3D leaderboard is inconsistent across different sources, with the version on GitHub still unreleased, while the one on HF Spaces is already available\footnote{\url{https://github.com/3DTopia/GPTEval3D/issues/3}}\\ 

\subsection{Missing Entity Smell}
This smell refers to the complete absence of expected entities on the leaderboard. Missing entities can cause concerns about data integrity, decreased user engagement, and hindered research progress. \\
\hrule \noindent \\
\emph{Component}: Benchmark Metric \\
\emph{Example}: The LLM Benchmarker Suite leaderboard lacks documentation detailing the metrics used for each benchmark dataset\footnote{\url{https://github.com/FormulaMonks/llm-benchmarker-suite/issues/9}}. \\
\hrule \noindent \\
\emph{Component}: Benchmark Protocol \\
\emph{Example}: The evaluation \& training settings were missing in the M3DBench leaderboard\footnote{\url{https://github.com/OpenM3D/M3DBench/issues/2}}. \\
\hrule \noindent \\
\emph{Component}: Evaluation Raw Dataset \\
\emph{Example}: The ``CHIP-CDEE'' dataset is missing from the MedBench leaderboard\footnote{\url{https://github.com/open-compass/opencompass/discussions/785}}. \\
\hrule \noindent \\
\emph{Component}: Evaluation Score\\
\emph{Example}: The GAOKAO-Bench leaderboard lacks evaluation scores for the ``fill-in-the-blank questions'' and ``open-ended questions'' tasks\footnote{\url{https://github.com/OpenLMLab/GAOKAO-Bench/issues/25}}. \\
\hrule \noindent \\
\emph{Component}: Model \\
\emph{Example}: The LawBench leaderboard's model links are all redirected to ``There is nothing here!''\footnote{\url{https://github.com/open-compass/LawBench/issues/7}} \\
\hrule \noindent \\
\emph{Component}: Ranking Dataframe \\
\emph{Example}: We identified $80$ missing PWC leaderboards, such as the ``Question Answering on MNIST'' leaderboard\footnote{\url{https://github.com/paperswithcode/sota-extractor/issues/105}}. \\
\hrule \noindent \\
\emph{Component}: Submission Channel/Protocol\\
\emph{Example}: The submission channels were missing from the Open ITA LLM Leaderboard\footnote{\url{https://huggingface.co/spaces/mii-llm/open_ita_llm_leaderboard/discussions/4}}. \\

\subsection{Redundant Entity Smell} 
This smell refers to the unnecessary repetition or unused presence of the same entities on the leaderboard. Redundant entities contribute to data clutter, increase the likelihood of analysis errors, and impose additional maintenance overhead. \\ 
\hrule \noindent \\
\emph{Component}: Benchmark Task \\
\emph{Example}: There are six scenarios that remain unused in the HELM Classic leaderboard\footnote{\url{https://github.com/stanford-crfm/helm/issues/2062}}. \\
\hrule \noindent \\
\emph{Component}: Evaluation Score\\
\emph{Example}: There are two identical evaluation records in the ``Memex Question Answering on MemexQA'' leaderboard\footnote{\url{https://github.com/paperswithcode/sota-extractor/issues/41}}. \\
\hrule \noindent \\
\emph{Component}: Model\\
\emph{Example}: There were two listed models with identical names in the KoLA leaderboard\footnote{\url{https://github.com/THU-KEG/KoLA/issues/6}}. \\
\hrule \noindent \\
\emph{Component}: Ranking Dataframe \\
\emph{Example}: There were dozens of duplicate ranking dataframes in the KoLA leaderboard\footnote{\url{https://github.com/THU-KEG/KoLA/issues/2}}. \\

\subsection{Unresponsive Entity Smell} 
This smell refers to entities on the leaderboard that are accessible online but do not respond to user interactions due to technical issues. Unresponsive entities can cause several problems, including user frustration, reduced credibility, and inefficiency of management.\\
\hrule \noindent \\
\emph{Component}:Evaluator \\
\emph{Example}: There are consistent runtime errors whenever any input example is provided to the model in the Multi-Modality Arena\footnote{\url{https://github.com/OpenGVLab/Multi-Modality-Arena/issues/26}}. \\
\hrule \noindent \\
\emph{Component}: Ranking Dataframe \\
\emph{Example}: We encountered $135$ HF leaderboards experiencing runtime errors, such as the GlitchBench leaderboard\footnote{\url{https://huggingface.co/spaces/glitchbench/Leaderboard/discussions/3}}. \\
\hrule \noindent \\
\emph{Component}: Submission Channel/Protocol \\
\emph{Example}: The KoLA leaderboard submission portal was unresponsive to user requests\footnote{\url{https://github.com/THU-KEG/KoLA/issues/17}}. \\

We find \textbf{``confusing entity'' to be the smell that occurs across the most ($8$) number of leaderboard components. This is closely followed by the smells of ``mismatched entity'' ($7$) and ``missing entity'' ($7$).} On the other hand, we find that \textbf{the ranking dataframe component suffers from all eight types of leaderboard smell. This is closely followed by evaluation record ($6$) and model ($5$).} 

Furthermore, Table~\ref{tab:workflow smell} provides a comprehensive mapping of the presence of various smell types in relation to workflow patterns. Our findings reveal that $P_4$ (Pointwise Model Evaluation) is associated with all eight types of leaderboard smells, followed by $P_1$ (External Evaluation Integration) and $P_2$ (Model Output Evaluation) with both six types of smells. We conjecture that $P_4$'s results stem from this pattern being the most commonly adopted ($89.98\%$, $428/481$) for non-PWC leaderboards, increasing the likelihood of encountering all smell types.On the other hand, $P_3$ (Direct Model Evaluation) has the least number ($2$) of associated smells, though this may be partly due to its relatively low occurrence ($0.38\%$) among the collected leaderboards. Additionally, we observed that the ``missing entity'' smell is the most prevalent, appearing across all five identified workflow patterns. Other common smells include ``confusing entity'', ``inaccessible entity'', and ``unresponsive entity'', which are found in four workflow patterns each. In contrast, the ``redundant entity'' smell appears the least frequently, with only one workflow pattern involved.

\begin{table}[t]
\centering
\caption{The presence of leaderboard smells across FM leaderboards following specific workflow patterns.}
\label{tab:workflow smell}
\begin{NiceTabular}{l|lllll}
& \multicolumn{5}{c}{\textbf{Workflow Pattern}} \\ 
\cline{2-6}
\textbf{Smell Type} & $P_1$ & $P_2$ & $P_3$ & $P_4$ & $P_5$ \\ \midrule
Confusing Entity & \checkmark & \checkmark & \checkmark & \checkmark &  \\ \midrule
Deprecated Entity &  & \checkmark &  & \checkmark &  \\ \midrule
Inaccessible Entity & \checkmark & \checkmark &  & \checkmark & \checkmark \\ \midrule
Misdisplayed Entity & \checkmark &  &  & \checkmark &  \\ \midrule
Mismatched Entity & \checkmark & \checkmark &  & \checkmark &  \\ \midrule
Missing Entity & \checkmark & \checkmark & \checkmark & \checkmark & \checkmark \\ \midrule
Redundant Entity &  &  &  & \checkmark &  \\ \midrule
Unresponsive Entity & \checkmark & \checkmark &  & \checkmark & \checkmark \\
\end{NiceTabular}
\end{table}

\begin{mybox}{$RQ_2$ Summary}
\begin{itemize}[leftmargin=*]
    \item We identify eight types of leaderboard smells across nine distinct components, revealing common issues affecting leaderboards' functionality, usability, fairness, scalability, or sustainability.
    \item Among the domain model components, the ``confusing entity'' smell is the most prevalent, followed by the ``mismatched entity'' and ``missing entity'' smells. The ranking dataframe is the most affected, followed by the evaluation score and the model.
    \item We find that the workflow pattern $P_4$ (Pointwise Model Evaluation) is associated with the most leaderboard smells, followed by $P_2$ (Model Output Evaluation) and $P_1$ (External Evaluation Integration). Also, the ``missing entity'' smell is the most prevalent across all workflow patterns, followed by ``confusing entity'', ``inaccessible entity'', and ``unresponsive entity''.
\end{itemize}
\end{mybox}
\section{Implications}
\label{sec:implications}

\emph{LBOps as a Discipline}: By positioning Leaderboard Operations (LBOps) as a distinct discipline, our study paves the way for establishing future best practices that enhance transparency and sustainability across ML evaluations. By formalizing LBOps workflows, researchers can establish consistent and reliable documentation across various ML leaderboards, for instance in the form of ``leaderboard cards''. Inspired by \href{https://huggingface.co/docs/huggingface_hub/package_reference/cards}{repository cards}, leaderboard cards could serve as universal standards to improve the quality and reliability of leaderboards. Emerging tools, such as the \href{https://huggingface.co/spaces/demo-leaderboard-backend/leaderboard}{Demo leaderboard} and \href{https://huggingface.co/spaces/freddyaboulton/gradio_leaderboard}{gradio\_leaderboard} are already streamlining the prototyping and deployment process for leaderboards, helping operators to maintain accurate and up-to-date information. 

\emph{Collaborative Evolution in Leaderboard Development}: LBOps brings together diverse stakeholders—data scientists, ML engineers, and software developers—who typically work in isolation. Our research uncovers initial traces of collaborations among leaderboard developers. Platforms including \href{https://huggingface.co/spaces/GTBench/GTBench}{GTBench} openly borrow templates from successful leaderboards such as the \href{https://huggingface.co/spaces/open-llm-leaderboard/open_llm_leaderboard}{Open LLM Leaderboard}, which highlights a growing culture of template reuse and best practices. Other leaderboards, including \href{https://huggingface.co/spaces/TTS-AGI/TTS-Arena}{TTS Arena}, \href{https://huggingface.co/spaces/rstless-research/italian_open_llm_leaderboard}{Italian Open LLM Leaderboard}, and \href{https://huggingface.co/spaces/Bowieee/StructEval_leaderboard}{StructEval}, follow similar adaptation paths, suggesting a dynamic evolution within the leaderboard ecosystem. Further investigation of this shared evolution could provide deeper insight for practitioners and researchers on how leaderboards influence each other and drive continuous improvements in LBOps.

\emph{Leaderboard Quality Assurance with Bill of Materials}: The prevalence of leaderboard smells—issues that compromise the evaluation and ranking of models—underscores the need for stricter quality control. Inspired by the AI Bill of Materials~\cite{xia2023trust} (AIBOM), we propose the creation of a ``Leaderboard Bill of Materials'' (LBOM) to increase transparency. LBOM would act as a formal machine-readable inventory that records every step of the leaderboard process, ensuring compliance with guidelines and making the supply chain of the models evaluated and their ranking (\eg, the model versions being evaluated, the benchmark versions being used) visible. Such an initiative would strengthen trust in leaderboards and create a more robust and transparent framework to verify the integrity of model evaluations. 

\emph{Encouraging Community Engagement for Leaderboards}: Our analysis reveals a significant lack of community interaction for leaderboards, particularly those hosted on PWC and independent platforms, which often limit communication to operators' email addresses. This restrictive approach curtails opportunities for meaningful feedback, collaboration, and knowledge sharing. To address this, we recommend establishing dedicated discussion forums tailored for FM leaderboards, leveraging lightweight platforms, such as Discord or Slack. These forums would facilitate direct communication among users, developers, and maintainers, enabling the exchange of feedback, the sharing of best practices, and the early detection of leaderboard smells. By fostering continuous dialogue, they can improve the overall quality of leaderboards, encourage active stakeholder participation, and ensure timely identification and resolution of emerging issues.

\emph{Need for a Comprehensive Leaderboard Comparison}: Despite initial efforts to create meta-leaderboards, such as the \href{https://huggingface.co/spaces/mrfakename/open-leaderboards-leaderboard}{Open Leaderboards Leaderboard}, there remains a critical gap—a systematic framework to evaluate and rank FM leaderboards. As the number of leaderboards grows, so does the challenge of selecting the right one for a specific need. Inconsistencies across leaderboards in performance metrics or model results further complicate this process, exacerbated by the prevalence of leaderboard smells that can undermine trust in their validity. To address these concerns, we advocate for the development of a comprehensive evaluation framework, similar to traditional software quality assessments. This framework should assess both the quality and performance attributes of the leaderboards. Such a system would provide stakeholders with the tools to make informed decisions, ensuring that the leaderboards they rely on are consistent, transparent, and reliable.
\section{Threats to Validity}
\label{sec:threats}

\subsection{Conclusion Validity} 

In cases where our efforts to establish communication through emails or issue trackers go unanswered, we rely on evidence-based deduction by examining available documentation, publications, and community resources to gather the necessary information. Although these analyses may introduce confirmation bias, the diverse expertise of the authors helps mitigate potential biases. The first author has practical experience in developing HF leaderboards, while the other authors are experienced SE researchers. Multiple authors are involved in the analysis process and conflicts are resolved using the negotiated agreement technique~\cite{campbell2013coding}. In cases where evidence was insufficient for further verification, particularly for leaderboards hosted on independent websites, we designated the relevant information as ``unknown'' to ensure transparency and maintain the reliability of our findings. 

\subsection{Construct Validity} 

In our study, we categorize workflow patterns based on the actions of key stakeholders—namely leaderboard operators, external contributors, and independent judges—and the evaluation methods used, such as pointwise and pairwise comparisons. This approach provides a structured framework for representing LBOps. However, we acknowledge that alternative categorization schemes could offer additional insight. For example, workflows could be categorized based on evaluator type, such as human-based, AI-based, or reference-based evaluations. Reference-based evaluations involve predefined, objective benchmarks, while human- and AI-based evaluations depend on subjective judgments or model-based assessments. Nevertheless, our domain model remains flexible, allowing for the integration of these alternative categorizations and perspectives within LBOps. 

Not all of our proposed smell reports received responses from operators, which may indicate potential false positives in our coding approach. However, the resolution rate for the reported smells improved from $35.11\%$ to $43.74\%$ over a four-month period, suggesting that many initially unresolved issues were eventually confirmed or addressed. Interestingly, some reports that were initially dismissed as non-issues were later acknowledged\footnote{\url{https://github.com/paperswithcode/paperswithcode-client/issues/24}} or resolved without explicit attribution\footnote{\url{https://github.com/open-compass/LawBench/issues/7}}, highlighting the evolving nature of issue recognition~\cite{zimmermann2010makes,bettenburg2008makes}.

Our analysis relies primarily on static documentation and commit histories, which may provide only a partial view of LBOps in real-world scenarios. This limitation makes it difficult to detect certain cases, such as significant delays in evaluations caused by a high volume of submission requests or the adoption of more comprehensive benchmarks. Consequently, the reported prevalence of smells and their distribution across workflow patterns may represent a lower bound. To address these issues, future work should consider incorporating dynamic, real-time data collection methods to provide a more comprehensive understanding of leaderboard dynamics.

To address these limitations and gain practical insight into the challenges of LBOps, we consult a prominent operator of \href{https://www.superclueai.com}{SuperCLUE}, taking advantage of his extensive experience in developing ML benchmarks and managing associated leaderboards. On \interviewdate, we present our draft in detail and seek feedback on our findings. The operator confirms that no major leaderboard smells are overlooked, except for performance-related issues such as benchmark leakage—a phenomenon where models gain an unfair advantage by memorizing data inadvertently included in the training set. While our observations indicate that benchmark leakage primarily affects leaderboards employing workflow patterns $P_2$ and $P_4$, we fail to detect these issues in our collected leaderboards. This is likely because benchmark leakage often occurs outside LBOps and cannot be directly observed through manual analysis. Addressing this challenge is inherently difficult for most leaderboard operators, as it requires robust detection mechanisms while maintaining a balance between transparency and usability.

Furthermore, the lead operator emphasizes that leaderboard workflows are highly context-dependent and are influenced by the publisher's specific needs and resource constraints. For example, while open-sourcing and enabling user-submitted models in reproducible Docker environments could enhance scalability and flexibility, such approaches are often impractical due to resource limitations. He also highlights the inevitability of issues like failed tests or outdated documentation, especially on leaderboards with declining maintenance—challenges that align with our identified leaderboard smells. Additionally, he notes that as leaderboards evolve, gaps in documentation and maintenance inevitably emerge over time. These insights highlight the importance of proactive management and robust support systems to ensure the long-term success of leaderboards. However, we acknowledge that feedback from a single leaderboard expert, while valuable, is inherently limited in scope. Gathering feedback from multiple leaderboard operators is an important direction for future research to systematically verify the completeness and generalizability of our observations.

\subsection{External Validity}

In our study, we exclude anonymous evaluations, such as those of \href{https://www.kaggle.com/competitions}{Kaggle Competitions}, as their anonymity prevents us from verifying whether they include FM evaluations. Despite our extensive multi-source analysis of $1,045$ FM leaderboards, this limitation may result in an incomplete or potentially skewed view of FM evaluations. Additionally, many leaderboards actively add, remove, and modify metrics, datasets, and evaluation frameworks over time. These changes may not align with our collected data, and consistently capturing them requires continuous monitoring—something a static collection method cannot fully achieve. To identify potential leaderboards we might have missed, we expand our search to online platforms, including Google, X, YouTube, and Google Scholar. However, these platforms did not reveal any new leaderboards, as our existing search heuristics had already captured all relevant results. Despite these efforts, we acknowledge the inherent limitations of our approach in fully capturing the rapidly evolving landscape of FM leaderboard practices.

While our study's focus is on FM leaderboards, we recognize the existence of various other types of ML leaderboards that evaluate and compare artifacts other than models, such as databases (\eg, \href{https://zilliz.com/vector-database-benchmark-tool}{VectorDBBench}), datasets (\eg, \href{https://www.datacomp.ai}{DataComp}), method (\eg, \href{https://github.com/wenhao728/awesome-diffusion-v2v/blob/main/doc/leaderboard.md}{V2VBench}), metrics (\eg, \href{https://github.com/yuh-zha/AlignScore}{AlignScore}), papers (\eg, \href{https://huggingface.co/spaces/ameerazam08/Paper-LeaderBoard}{Papers Leaderboard}), and even leaderboards themselves (\eg, \href{https://huggingface.co/spaces/mrfakename/open-leaderboards-leaderboard}{Open Leaderboards Leaderboard}). Additionally, we exclude ML leaderboards that host smaller, non-foundation models. Expanding our research to include these ML leaderboards could offer deeper insight into the various features and processes within the broader LBOps framework. 


\subsection{Internal Validity} 

The identification of workflow patterns and smells in LBOps may be influenced by human biases, including researchers' personal perspectives, experiences, or emotional states during coding, which could impact the accuracy and impartiality of our findings. To address this, we conduct weekly meetings among the authors using the negotiated agreement approach~\cite{campbell2013coding}. This method facilitates consensus on code definitions and fosters a shared understanding of the coding criteria, thereby enhancing the reliability and objectivity of our analysis.
\section{Conclusion}
\label{sec:conclusion}

In this study, we explore the inherent features and pitfalls in LBOps from the perspective of leaderboard users by examining up to $1,045$ FM leaderboards. First, we define the discipline of ``leaderboard operations'' (LBOps), which encompasses five distinct workflow patterns, each catering to different FM evaluation and ranking requirements. Simultaneously, we derive a domain model to encapsulate all concepts involved in the workflow patterns. Then, we identify eight types of ``leaderboard smells'', deteriorating the sustainability and trustworthiness of FM leaderboards. While our study focuses primarily on FM leaderboards, we believe that our findings can also be extended to leaderboards hosting comparisons of smaller models. On the one hand, leaderboard operators can use our insights to improve their LBOps practices in FM comparison and selection. On the other hand, SE teams, as prominent FM users~\cite{feng2024prompting,nam2024using}, can leverage our findings to make more informed decisions when selecting the most appropriate leaderboards for their needs. 
\section*{Acknowledgement}

We sincerely thank Mr. Liang Xu\footnote{\url{https://github.com/brightmart}\label{footnote:Liang Xu}}, the lead operator of \href{https://www.superclueai.com}{SuperCLUE}, for his invaluable insights and feedback.

\bibliographystyle{spmpsci}
\bibliography{references}

\end{document}